\documentclass[review,5p,twocolumn]{elsarticle}

\usepackage{hyperref}
\journal{Pervasive and Mobile Computing}

\usepackage{graphicx} 
\usepackage[ruled,linesnumbered,longend]{algorithm2e} 

\usepackage{caption}
\usepackage{subcaption}
\usepackage{etoolbox}
\usepackage{epstopdf} 
\usepackage{xcolor}

\usepackage{amsthm}

\newtheorem{theorem}{Theorem}
\newtheorem{lemma}{Lemma}

\newtheorem{definition}{Definition}

\begin{document}

\begin{frontmatter}

\title{An Algorithmic Study in the Vector Model for Wireless Power Transfer Maximization\tnoteref{t1}}
\tnotetext[t1]{A preliminary version of this paper appeared in \cite{2017icdcn}.}
%\tnotetext[mytitlenote]{Fully documented templates are available in the elsarticle package on \href{http://www.ctan.org/tex-archive/macros/latex/contrib/elsarticle}{CTAN}.}

%% Group authors per affiliation:
%\author{Theofanis. P. Raptis\fnref{myfootnote}}
\author[add2,add3]{Ioannis Katsidimas\corref{cor1}}
   \ead{ikatsidima@ceid.upatras.gr}
   \author[add2,add3]{Sotiris Nikoletseas}
   \ead{nikole@cti.gr}
   \author[add1]{Theofanis P. Raptis}
   \ead{theofanis.raptis@iit.cnr.it}
   \author[add2,add3]{Christoforos Raptopoulos}
   \ead{raptopox@ceid.upatras.gr}
 
   \cortext[cor1]{Please address correspondence to I. Katsidimas}
   \address[add1]{Institute of Informatics and Telematics, National Research Council, Pisa, Italy}
   \address[add2]{Department of Computer Engineering and Informatics, University of Patras, Patras, Greece}
   \address[add3]{Computer Technology Institute and Press Diophantus, Patras, Greece}
%\fntext[myfootnote]{Since 1880.}

%% or include affiliations in footnotes:
%\author[mymainaddress,mysecondaryaddress]{Elsevier Inc}
%\ead[url]{www.elsevier.com}

%\author[mysecondaryaddress]{Global Customer Service\corref{mycorrespondingauthor}}
%\cortext[mycorrespondingauthor]{Corresponding author}
%\ead{support@elsevier.com}

%\address[mymainaddress]{1600 John F Kennedy Boulevard, Philadelphia}
%\address[mysecondaryaddress]{360 Park Avenue South, New York}

\begin{abstract}
Rapid technological advances in the domain of Wireless Power Transfer (WPT) pave the way for novel methods for power management in systems of wireless devices and recent research works have already started considering algorithmic solutions for tackling emerging problems. However, many of those works are limited by the system modelling, and more specifically \textit{the one-dimensional abstraction} suggested by Friis formula for the power received by one antenna under idealized conditions given another antenna some distance away.\newline\indent
Different to those works, we \textit{use a model} which arises naturally from fundamental properties of the \textit{superposition of energy fields}. This model has been shown to be more realistic than other one-dimensional models that have been used in the past and can capture superadditive and cancellation effects. Under this model, we define \textit{two new interesting problems} for configuring the wireless power transmitters so as to maximize the total power in the system and we prove that the first problem can be solved in polynomial time. We present a \textit{distributed solution} that runs in pseudo-polynomial time and uses various knowledge levels and we provide theoretical performance guarantees. Finally, we design \textit{three heuristics} for the second problem and evaluate them via simulations.
\end{abstract}

\begin{keyword}
Wireless Power Transfer\sep Distributed Algorithms\sep Wireless Systems
\end{keyword}

\end{frontmatter}

%\linenumbers

\section{Introduction}
\label{sec:1}
Wireless Power Transfer (WPT) has recently become a commercially viable option in various wireless systems due to the reliability of continuous power supply and the convenience provided by the fact that no static (wired) network connections are needed between the devices. The efficiency of the various technological alternatives is increasing every year. Current fast-charging protocols achieve up to 84\% efficiency for wireless power transfer up to distances of 15 meters \cite{powercast}, while at the same time keeping thermal dissipation significantly low \cite{ti}.
%Currently, several solutions can achieve up to 84\% efficiency with multiple fast-charging protocol support and significantly low thermal dissipation \cite{ti}, able to transmit up to distances of 15 meters \cite{powercast}.
%The Wireless Power Consortium \cite{wpc} is a group of leading manufacturers founded in 2008 in a wide range of industries that understand the untapped potential of wireless charging. Its members include leading makers of mobile phones, consumer electronics, batteries, semiconductors, components, wireless power technology and infrastructure. 
A WPT enabled system consists of several wireless transmitter and receiver devices. A \emph{wireless transmitter (charger)} is a device that has a dedicated power source with significant power supply and can transfer power wirelessly to receivers. A \emph{receiver (node)} is a device that is powered by harvesting the radio frequency energy from the chargers. A receiver is usually an electronic device that is needed to perform a specific task in the wireless system, for example a sensor mote in a wireless sensor network. Systems of wireless devices have to operate under increasing demands of power in order to sustain various computational and communication tasks. For this reason, the efficient and distributed cooperation of the transmitters and receivers towards achieving an effective power allocation in the system is a crucial task. %An important goal in the design and efficient implementation of large wireless systems is to save energy and keep the network functional for as long as possible. This can be achieved by using WPT as a power exchange enabling technology and applying interaction protocols among the agents which guarantee that the available power in the network can be eventually distributed efficiently.

While considerable research efforts have been invested into power management in wireless systems, most of the models studied in the WPT applications literature neglect key features of electromagnetic fields which have been proposed by works in the WPT technology design literature. For example, physical methods, based on the electromagnetic theory and on numerical simulations, are adopted to predict the actual electromagnetic energy in selected locations accounting for the effective radio channel describing the environment. These methods use rigorous models of the antennas and of the transmitter and receivers and also study non conventional signal excitations to enhance the transfer efficiency. Some indicative works are \cite{7054689} and \cite{6916985}.

As a result, such models are unable to explain various phenomena occurring in real applications (e.g., cancellation and superadditive charging effects; see Section \ref{sec:model}), yielding some algorithmic solutions impractical. 

\textbf{Our contribution.} In view of the above, our contribution in this chapter is the following:
\begin{itemize}
\item We algorithmically study a more realistic model for WPT at far-field region in wireless systems, which was nicely initiated in \cite{7127689}. In particular, this is \textit{a ``vector'' model} that takes into account \textit{the superposition} of electromagnetic fields created by independent wireless transmitters, as well as fundamental properties of the superposition of waves from physics. To the best of our knowledge, this is the first algorithmic study of a vector model for WPT.

%To the best of our knowledge, this is the first study that uses a detailed, realistic model in an algorithmic context for wireless systems. The model takes into account the superposition of electromagnetic fields created by independent wireless transmitters, as well as fundamental properties of the superposition of waves from physics.
\item We define \textit{two new computational problems} for the efficient utilization of power resources in a wireless system consisting of a family of transmitters ${\cal C}$ and a family of receivers ${\cal R}$. In particular, we first consider the problem \texttt{MAX-POWER} of finding a configuration (set of operation levels) for transmitters that \textit{maximizes the total power} received by ${\cal R}$. Second, we consider the problem \texttt{MAX-$k$MIN-GUARANTEE} of finding a configuration that maximizes \textit{the minimum cumulative power \footnote{The cumulative power for a set of nodes is the aggregate received power by the nodes of a set from all the operational chargers.} among all the sets of $k$ receivers}.

\item We formulate \texttt{MAX-POWER} as a \textit{quadratic program} and we prove that we can find an optimal solution efficiently by presenting \textit{a family of distributed algorithms} using different levels of knowledge of the system. We prove that these algorithms run in pseudopolynomial time, but we show that they are quite faster in practice. 

\item We design and evaluate \textit{three efficient heuristics} for \texttt{MAX-$k$MIN-GUARANTEE} that provide good approximations of the optimal solution. The first heuristic is a generalization of our algorithmic solution to \texttt{MAX-POWER}, while the second samples a few representative $k$-sets of receivers and then solves the problem considering only those sets. Finally, our third heuristic is a hybrid of the previous two ideas and we show that it outperforms both in typical deployments of transmitters and receivers on the plane. 

\end{itemize}
%A preliminary version of this work \cite{Katsidimas2017} was appeared in the proceedings of the 18th International Conference on Distributed Computing and Networking (ICDCN 2017).

\section{Related Work}
\label{sec:2}

Wireless Power Transfer methods in large scale networked systems have attracted much attention from researchers worldwide. The reader can find comprehensive reviews of the relevant literature in \cite{7327131, 6951347, 6590061}, and  \cite{7096295}. Also, the book \cite{wpt-book}, is the first systematic exposition on the domain of wireless energy transfer in ad hoc communication networks. Several works study applications in sensor networks \cite{Angelopoulos2014113,6566993,6153401} and wireless distributed systems \cite{Madhja201589,raptopox}, UAVs \cite{6225205,Mittleider2016}. Different to all those works, in this article we investigate a static setting, where transmitter and receiver locations are predefined and stationary, and there are no mobile elements in the system.

There have been some works on closely related themes to this article, which investigate different aspects of the Wireless Power Transfer process. In \cite{7218622}, given a set of candidate locations for placing chargers, the authors provide a charger placement and a corresponding power allocation to maximize the charging quality, subject to a power budget. In a recent paper \cite{7164906}, a subset of the authors of this article study the Low Radiation Efficient Charging Problem, in which we optimize the amount of useful energy transferred from chargers to nodes (under constraints on the maximum level of imposed electromagnetic radiation). In a similar setting, the authors in \cite{6888920} consider the Safe Charging with Adjustable Power (SCAPE) problem for adjusting the power of chargers to maximize the charging utility of devices, while assuring that electromagnetic intensity at any location in the field does not exceed a given threshold. It is worth noting that, even though all the above works nicely demonstrate the gains of carefully distributing the power in a wireless setting, they use one-dimensional models, which fall short of capturing various intricate aspects of WPT.

Wireless transfer of energy through directed radio frequency waves has the potential to realise perennially operating sensor nodes by replenishing the energy contained in the limited on-board battery. However, the high power energy transfer from energy transmitters interferes with data communication, limiting the coexistence of these functions. In \cite{7037190}, the authors provide an experimental study to quantify the rate of charging, packet loss due to interference, and suitable ranges for charging and data communication of the energy transmitters. They also explore how the placement and relative distances of multiple energy transmitters affect the charging process, demonstrating constructive and destructive energy aggregation at the sensor nodes. Finally, the authors investigate the impact of the separation in frequency between data and energy transmissions, as well as among multiple concurrent energy transmissions. Their results aim at providing insights on radio frequency-based energy harvesting wireless sensor networks for enhanced protocol design and network planning.

Furthermore, in \cite{7127689}, the authors formulate the location-dependent power harvesting rates in generalised 2D and 3D placement of multiple (RF) energy transmitters for recharging the nodes of a wireless sensor network. In particular, they study the distributions of total available and harvested power over the entire network. They then provide closed matrix forms of harvestable power at any given point in space due to the action of concurrent energy transfer from multiple energy transmitters, explicitly considering constructive and destructive interference of the transmitted energy signals. The authors also analyse the performance of energy transfer in the network through power outage probability, interference, and harvested voltage as a function of the wireless energy received from the energy transmitters. The results reveal that the network wide received power and interference power from concurrent energy transfers exhibit Log-Normal distributions, and the harvested voltage over the network follows a Rayleigh distribution.

In \cite{5541663}, the authors investigate the impact of wireless charging technology on sensor network deployments and routing arrangements, formalise the deployment and routing problem, prove it as $\mathcal{NP}$-complete, develop heuristic algorithms to solve the problem, and evaluate the performance of the solutions through extensive simulations. 

In \cite{Gao2014395}, the authors introduce a scheme for improving the transmission power of nodes to bound end to end delay. They provide an algorithm for finding the minimal sleep latency from a node to a sink by increasing the minimal number of nodes whose transmission power improved. For bounding the end to end delay from the source node to the sink, the authors propose an end to end delay maintenance solution and demonstrate its efficiency in providing end to end delay guarantees in rechargeable wireless sensor networks

Finally, in \cite{6566753,6782467}, through an experimental study, the authors first demonstrate how the placement, the chosen frequency, and number of the RF energy transmitters affect the sensor charging time. These studies are then used to design a MAC protocol called RF-MAC that optimises energy delivery to desirous sensor nodes on request.

\section{The charging model} \label{sec:model}

In a recent paper (\cite{7127689}), the authors considered a model for the superposition of electromagnetic fields created by independent wireless energy sources, which takes into account fundamental properties of the superposition of waves from physics. The model of \cite{7127689} goes beyond (in fact, it is a generalization of) the one-dimensional abstraction suggested by Friis' formula\footnote{Given two antennas, the ratio of power available at the input of the receiving antenna, $P_r$, to output power to the transmitting antenna, $P_t$, is given by $\frac{P_r}{P_t}=G_tG_r(\frac{\lambda}{4\pi R})^2$ where $G_t$ and $G_r$ are the antenna gains (with respect to an isotropic radiator) of the transmitting and receiving antennas respectively, $\lambda$ is the wavelength, and $R$ is the distance between the antennas.} for the power received by one antenna under idealized conditions given another antenna some distance away. In particular, the \emph{electric field} created by an energy transmitter (charger) $C$, operating at full capacity, at a receiver $R$ at distance $d = \mathrm{dist}(C, R)$ is a 2-dimesional vector given by 
\begin{equation} \label{eq:vector-field}
\mathbf{E}(C, R) \stackrel{def}{=} \beta \cdot \frac{1}{d} \cdot e^{-j \frac{2\pi}{\lambda} d} = \beta \cdot \frac{1}{d} \cdot \left[
\begin{array}[c]{c}
	\cos{\left(\frac{2\pi}{\lambda} d \right)} \\
	\sin{\left(\frac{2\pi}{\lambda} d \right)}
\end{array} \right],
\end{equation}
where $\lambda$ depends on the frequency at which the transmitter operates, and $\beta$ is a constant that depends on the hardware of the transmitter and the environment.\footnote{In fact, the exact formula used in \cite{7127689} for the electric field is $\mathbf{E}(C, R) \stackrel{def}{=} \sqrt{\frac{Z_0 G_C P_C}{4\pi d^2}} \cdot e^{-j \frac{2\pi}{\lambda} d}$, where $Z_0$ is a physical constant indicating the wave-impendance of a plane wave in free space, $G_C$ is the gain and $P_C$ is the output power of the transmitter. In this chapter, without loss of generality of our algorithmic solutions, we assume that all wireless transmitters and receivers are identical, thus the aforementioned parameters are the same for each charger.} 

The main point of the charging model of \cite{7127689}, which also sets it apart from other (less realistic, but more tractable) models in the wireless charging literature, is that the total electric field created by a family of energy transmitters ${\cal C}$ at a receiver $R$ is the \emph{superposition (vector-sum)} of their individual electric fields, that is
\begin{equation} \label{eq:vector-additive}
\mathbf{E}({\cal C}, R) \stackrel{def}{=} \sum_{\mathbf{C} \in {\cal C}} \mathbf{E}(C, R).
\end{equation}
Furthermore, the total available \emph{power} at the receiver $R$ is given by 
\begin{equation}
P({\cal C}, R) = \gamma \cdot \|\mathbf{E}({\cal C}, R)\|^2,
\end{equation}
where $\|\cdot\|$ denotes the length (2-norm) of the vector. The constant $\gamma$ depends on the hardware of the transmitter, the hardware of the receiver and the RF-to-DC conversion efficiency. 

\begin{figure}[t]    
    \begin{subfigure}[b]{\columnwidth}
    \centering
        \includegraphics[width=0.7\textwidth]{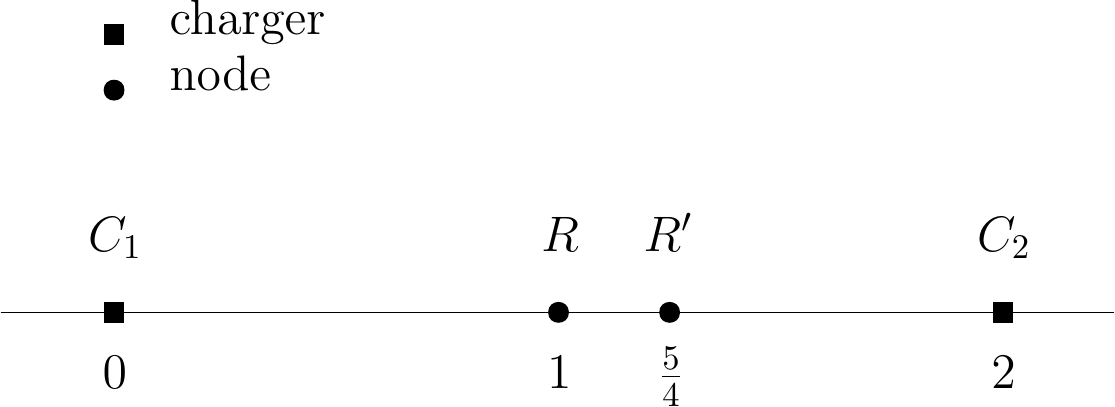}
        \captionsetup{justification=centering}
        \caption{Chargers' and nodes' placement on a straight line at points $(0, 0)$, $(0, 2)$ and $(0, 1)$, $(0, \frac{5}{4})$ respectively.}
        \label{fig:toydeployment}
    \end{subfigure}
    \centering
    \begin{subfigure}[b]{0.9\columnwidth}
        \includegraphics[width=\textwidth]{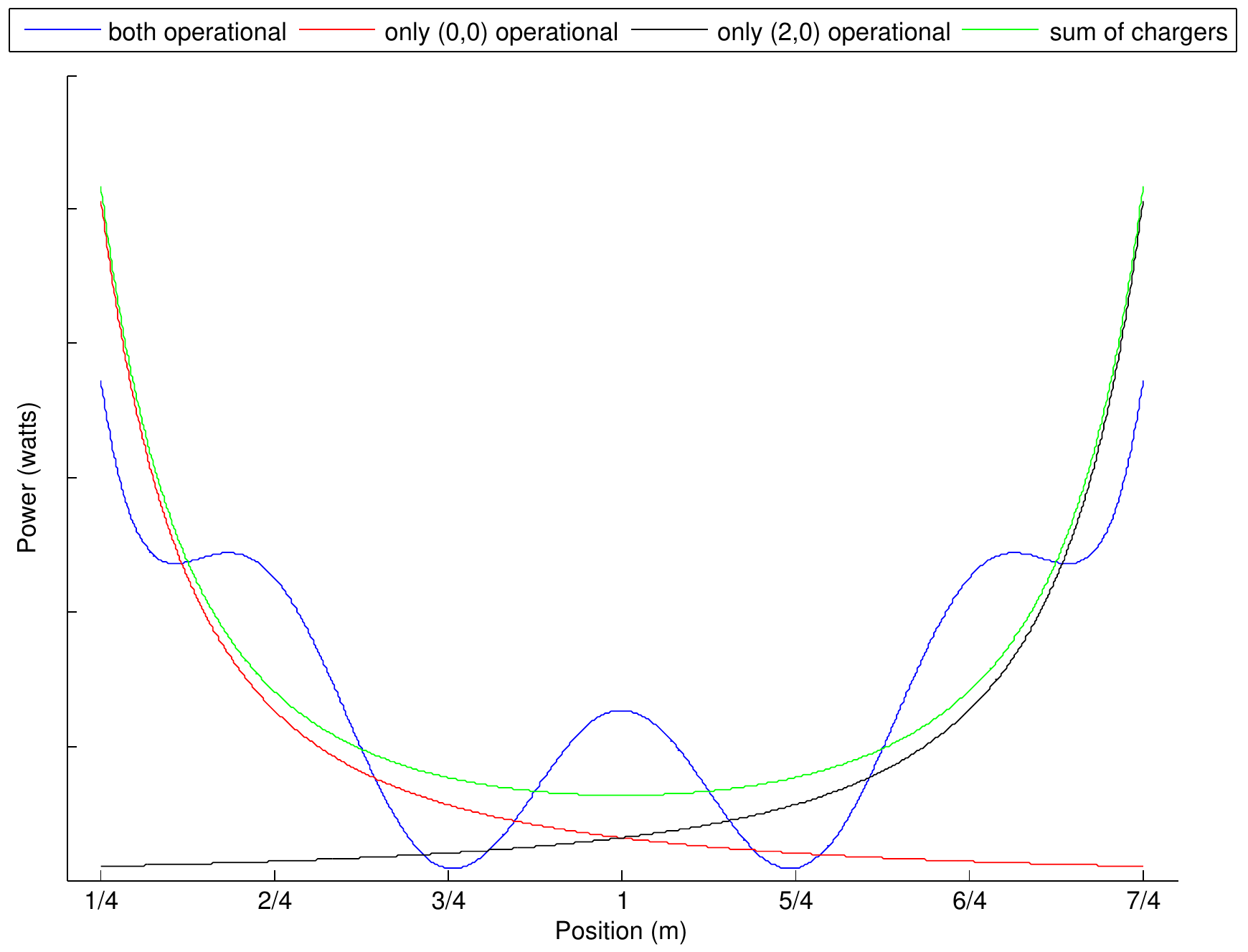}
        \caption{The power distribution between the two chargers. Different curves represent different operation levels of the chargers.}
        \label{fig:toyexample}
    \end{subfigure}
    ~ %add desired spacing between images, e. g. ~, \quad, \qquad, \hfill etc. 
    %(or a blank line to force the subfigure onto a new line)
    \caption{Example showing the superadditive and cancelation effects.}\label{fig:figtoyexample}
\end{figure}

It is worth noting that the above model arises naturally from fundamental properties of the superposition of energy fields and has been shown to be more realistic than other one-dimensional models that have been used in the past and can capture \emph{superadditive} and \emph{cancellation} effects \cite{6295618,7194761,7037190}. To fix ideas and to demystify the above definitions, we present the following fictitious example: Assume there are two transmitters $C_1$ and $C_2$ placed at points $(0, 0)$ and $(2, 0)$ in the 2-dimensional plane. First, consider a receiver $R$ placed at $(1, 0)$. Assume also, for the sake of clarity that all constants in the above model are set to 1, i.e. $\lambda = \beta = \gamma =1$. When only one of the two transmitters is operational, the power received by $R$ is $P(C_1, R) = P(C_2, R) = \|\mathbf{E}(C_1, R)\|^2 = \|\mathbf{E}(C_2, R)\|^2 = \left(\frac{1}{\mathrm{dist}(C_1, R)} \right)^2 =1$. On the other hand, if both transmitters are operational, the power received by $R$ is given by equation (\ref{eq:vector-additive}), that is 
\begin{displaymath}
P(\{C_1, C_2\}, R) = \|\mathbf{E}(C_1, R) + \mathbf{E}(C_2, R)\|^2.
\end{displaymath}
Furthermore, it is not hard to see that, since $R$ is equidistant from either $C_1$ or $C_2$, the vectors $\mathbf{E}(C_1, R)$ and $\mathbf{E}(C_2, R)$ point to the same direction. Therefore, $P(\{C_1, C_2\}, R) = 4 P(C_1, R) = 2 (P(C_1, R) + P(C_2, R)) = 4$. Notice then that the power received by $R$ when both transmitters are operational is larger than the sum of the powers it receives when only one of the transmitters is operational; this is the so-called superadditive effect and is visible in local maxima in the curve shown in Figure \ref{fig:toyexample}.

Second, consider a receiver $R'$ placed at $\left(\frac{5}{4}, 0 \right)$. Then by equation (\ref{eq:vector-field}), $\mathbf{E}(C_1, R') = \frac{4}{5} \cdot \left[
\begin{array}[c]{c}
	0 \\
	1
\end{array} \right]$, and also $\mathbf{E}(C_2, R') = \frac{4}{3} \cdot \left[
\begin{array}[c]{c}
	0 \\
	-1
\end{array} \right]$. By equation (\ref{eq:vector-additive}), the power received by $R'$ when both transmitters are operational is $P(\{C_1, C_2\}, R') = \left(\frac{8}{15}\right)^2 \approx 0.28$. Notice then that the power received by $R'$ when both transmitters are operational is much less than $\min\{P(C_1, R'), P(C_2, R')\} = \left(\frac{4}{5}\right)^2 \approx 0.64$; this is the so-called cancellation effect and is visible in local minima in the curve shown in Figure \ref{fig:toyexample}.

In view of our discussion in this chapter, it is worth noting that, even in the above toy example, it is non-trivial to provide a closed formula for the point in the line between the two chargers where the received power is maximized. 

\begin{figure}[t]
    \centering  
    \begin{subfigure}[b]{\columnwidth}
        \includegraphics[width=\textwidth]{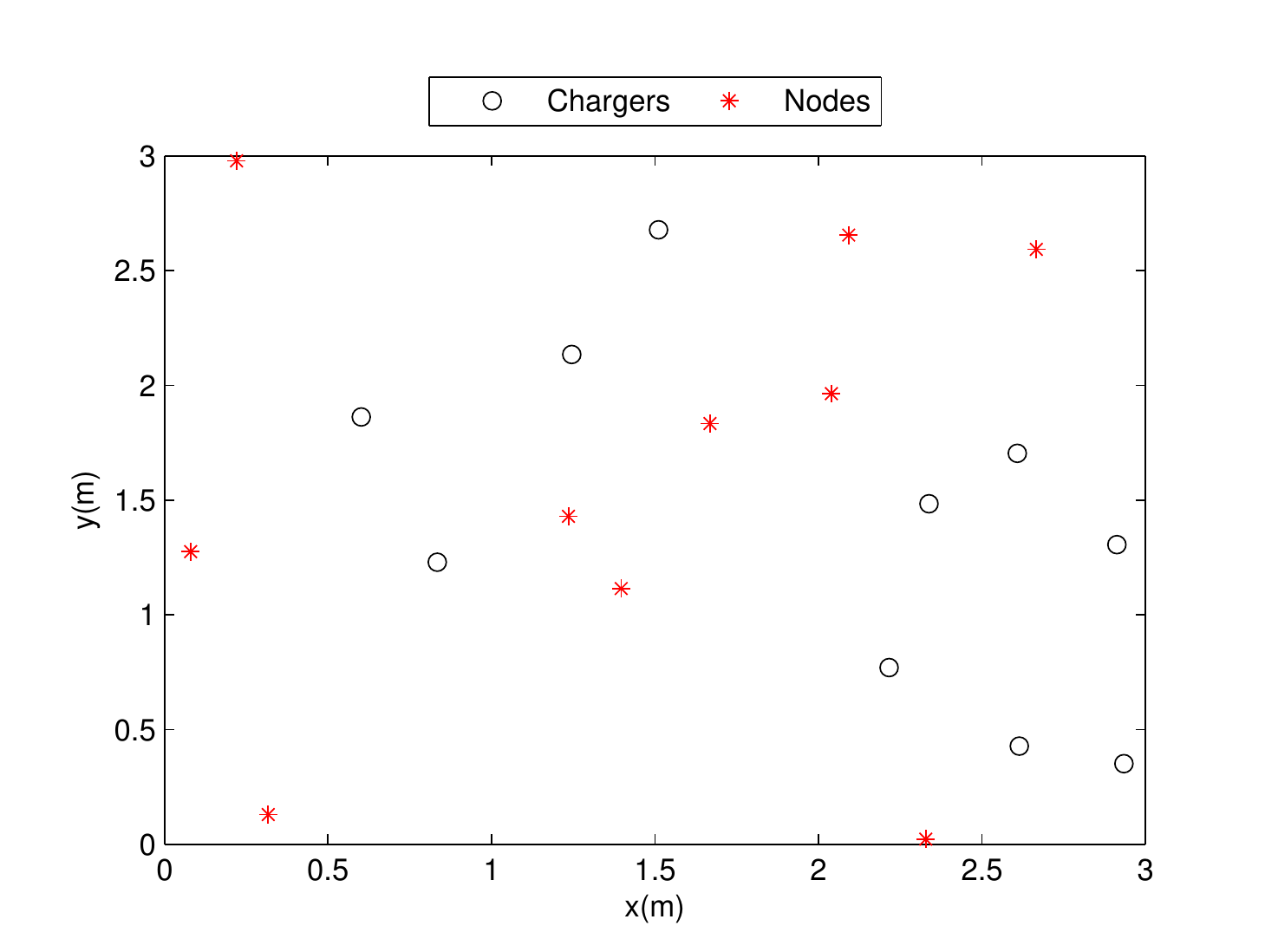}
        \captionsetup{justification=centering}
        %\caption{Power balance in general case}
        %\label{fig:pbgeneral}
    \end{subfigure}
    \caption{Chargers' and nodes' deployment for the simulation.}\label{fig:toyexampleplacement}
\end{figure}

We also conducted a simple simulation in a 3m $\times$ 3m plane with 10 chargers and 10 nodes, randomly deployed in the field (Fig. \ref{fig:toyexampleplacement}). The parameters of this simulation are the same with those mentioned at Section \ref{evaluation}. In this simulation we present the results of two different configurations. As the scalar model suggests, we turn on all the chargers to get the maximum cumulative received power which is valued to 0,217 Watts (the evaluation is done with respect to the vector model). On the other hand, a better configuration with respect to the vector model, turns on all the chargers except of the two at position $(2.2165,0.7708)$ and $(2.9343,0.3512)$. That configuration returns 0,230 Watts of cumulative received power in the network. Therefore, the claim of the toy example holds also for a more general case.

Finally, to further verify the vector model, we present the experiment, described in the technical report of \cite{7194761}. In this experiment, the distance between the node and charger 1 is fixed to $0.3m$ while the distance between the node and  charger 2 varies from $0.4m$ to $1.1m$ in increments of $0.1 m$. The results are presented in Table \ref{a}, where the second and third row record the received power when either charger 1 or charger 2 is turned on. The fourth row records the actual power measured when both chargers are turned on. The next two rows give the results for each of the two models we compare. Observe that the vector model approaches the actual measurements of row three better than the scalar model. This can be verified from the last two rows, which present the relative error of each model.

\begin{table*}
\centering
\begin{tabular}{|c|c|c|c|c|c|c|c|c|}
\hline
Harvesting power (mW)/Distance to charger 2 (m) & 0.4 & 0.5 & 0.6 & 0.7 & 0.8 & 0.9 & 1.0 & 1.1 \\ \hline
Charger 1 & 6.48 & 6.48 & 6.48 & 6.48 & 6.48 & 6.48 & 6.48 & 6.48 \\ \hline
Charger 2 & 3.21 & 1.92 & 1.96 & 0.09 & 0.71 & 0.33 & 0.35 & 0.06 \\ \hline
Charger 1 and Charger 2 & 3.57 & 2.05 & 3.47 & 7.78 & 2.11 & 1.32 & 9.04 & 6.04 \\ \hline
Scalar model (Friis) & 9.69 & 8.40 & 8.44 & 6.57 & 7.19 & 6.81 & 6.83 & 6.54 \\ \hline
Vector model & 6.21 & 5.11 & 8.19 & 6.58 & 5.77 & 6.62 & 6.74 & 6.42 \\ \hline
Relative error of scalar model (Friis) & 6.12 & 6.35 & 4.97 & -1.21 & 5.08 & 5.49 & -2.21 & 0.14 \\ \hline
Relative error of vector model & 2.64 & 3.06 & 4.72 & -1.2 & 3.66 & 5.3 & -2.3 & 0.02 \\ \hline
\end{tabular}
\caption{Charging power of two chargers (results from the technical report of \cite{7194761}).}
\label{a}
\end{table*}

\subsection{A basic assumption} \label{sec:assumptions}

A necessary condition that is required in Friis' formula, and by extension in the above model, is that the wireless transmitter and receiver need to be at distance at least $\lambda$ (not in the near-field region). In fact, for transmitters to receivers distances less than $\lambda$, more complex laws apply but we do not consider them here as they are beyond the algorithmic focus of the paper. A similar constraint also holds for the power received by receivers that are very close to each other. In particular, if receivers $R, R'$ are closer than $\frac{\lambda}{2\pi}$ apart, then the power received by each receiver no longer follows Friis' law. 
This was to be expected, since otherwise we could have wireless transmitters of bounded capabilities that could theoretically provide infinite power (e.g. by placing receivers arbitrarily close to each other and to the transmitter).  

In this chapter, we consider algorithmic problems related to power guarantees with respect to a fixed deployment of a family ${\cal C}$ of wireless transmitters and a family ${\cal R}$ of receivers. To avoid confusion, we will assume that any placement of chargers and receivers satisfies the above \emph{placement constraints}. In particular: (a) For each charger $C \in {\cal C}$ and receiver $R \in {\cal R}$, we have $\mathrm{dist}(C, R) \geq \lambda$ and (b) for any pair of receivers $R, R' \in {\cal R}$, we have $\mathrm{dist}(R, R') \geq \frac{\lambda}{2\pi}$.

We finally note that, since $\frac{\lambda}{2\pi}$ is usually smaller than a few centimeters, in practical situations, the above placement constraints will not be restrictive, as the non-trivial volume of any transmitting or receiving device guarantees that transmitters and chargers are far enough from each other.

% For figures use
%
%\begin{figure}[b]
%\sidecaption
%% Use the relevant command for your figure-insertion program
%% to insert the figure file.
%% For example, with the graphicx style use
%\includegraphics[scale=.65]{figure}
%%
%% If no graphics program available, insert a blank space i.e. use
%%\picplace{5cm}{2cm} % Give the correct figure height and width in cm
%%
%\caption{If the width of the figure is less than 7.8 cm use the \texttt{sidecapion} command to flush the caption on the left side of the page. If the figure is positioned at the top of the page, align the sidecaption with the top of the figure -- to achieve this you simply need to use the optional argument \texttt{[t]} with the \texttt{sidecaption} command}
%\label{fig:1}       % Give a unique label
%\end{figure}

\section{Problem definition}

Consider a system consisting of a family ${\cal C}$ of identical wireless chargers and a family ${\cal R}$ of identical wireless receivers (nodes). For each charger $C \in {\cal C}$, we denote by $\mathbf{x}_C \in [0, 1]$ a variable that determines the level of operation of $C$. We assume that the chargers are able to operate at a level of operation between $0\%$ and $100\%$. $x$ equal to 1 means that $C$ is fully operational (i.e. operates at $100\%$ capacity), while $x$ equal to 0 means that $C$ is non operational (operates at $0\%$). Clearly, we must have $x \in [0,1]$. In particular, generalizing equation (\ref{eq:vector-field}), the electric field vector created by $C$ at the location of a receiver $R$, when the former operates at level $\mathbf{x}_C$ is given by $\mathbf{x}_C \cdot \mathbf{E}(C, R) = \mathbf{x}_C \cdot \beta \cdot \frac{1}{\mathrm{dist}(C, R)} \cdot e^{-j \frac{2\pi}{\lambda} \mathrm{dist}(C, R)}$.  

We will refer to the vector $\mathbf{x} \in [0, 1]^{\cal T}$ as the \emph{configuration} of the chargers in the system. Slightly abusing notation, we will denote by ${\cal C}(\mathbf{x})$ a family of chargers that operate according to configuration $\mathbf{x}$.  

We initially consider the following problems where the locations of the wireless chargers and receivers are assumed to be known in advance.

\begin{definition}[\texttt{MAX-POWER}]
Given a family of chargers ${\cal C}$ and family of receivers ${\cal R}$ that satisfy the placement constraints of Subsection \ref{sec:assumptions}, find a configuration for the chargers that maximizes the total power to ${\cal R}$. That is, find $\mathbf{x}^*$ such that 
\begin{equation}
\mathbf{x}^* \in \arg\max_{\mathbf{x} \in [0, 1]^{\cal C}} P({\cal C}(\mathbf{x}), {\cal R}),
\end{equation}
where $P({\cal C}(\mathbf{x}), {\cal R}) = \sum_{R \in {\cal R}} P({\cal C}(\mathbf{x}), R)$.
\end{definition}

We will denote by ${{\cal R} \choose k}$ the family of all subsets of ${\cal R}$ containing $k$ nodes. In this chapter, we also study the following generalization of \texttt{MAX-POWER}, which finds a configuration that provides a minimum charging guarantee among all $k$-sets of nodes: 

\begin{definition}[\texttt{MAX-$k$MIN-GUARANTEE}]
Given a family of chargers ${\cal C}$ and a family of receivers ${\cal R}$ that satisfy the placement constraints of Subsection \ref{sec:assumptions}, find a configuration for the chargers that maximizes the minimum cumulative power among all subsets of ${\cal R}$ of size $k$. That is, find $\mathbf{x}^*$ such that
\begin{equation}
\mathbf{x}^* \in \arg\max_{\mathbf{x} \in [0, 1]^{\cal C}} \min_{A \in {{\cal R}\choose k}} P({\cal C}(\mathbf{x}), A),
\end{equation}
where $P({\cal C}(\mathbf{x}), A) = \sum_{R \in A} P({\cal C}(\mathbf{x}), R)$.
\end{definition}

\section{Maximum Total Power}

In this section we present an efficient algorithm for \texttt{MAX-POWER}. For simplicity, consider a family of wireless chargers ${\cal C} = \{C_1, \ldots, C_m\}$, where $m = |{\cal C}|$, and a family of receivers ${\cal R} = \{R_1, \ldots, R_n\}$, where $n = |{\cal R}|$. Let also $\mathbf{x} \in [0, 1]^m$ be the configuration of the chargers, where $\mathbf{x}_j$ is the level of operation of charger $C_j, j \in [m]$. 

We first show that the \texttt{MAX-POWER} problem can be expressed as a quadratic program. To this end, for each $R \in {\cal R}$, define $\mathbf{Q}^{(R)}$ be a $2 \times m$ matrix whose $j$-th column is the 2-dimensional vector of the electric field created from $C_j$ at $R$, i.e. $\mathbf{Q}^{(R)}_{:, j} = \sqrt{\gamma} \cdot \mathbf{E}(C_j, R)$, for each $j \in [m]$. Notice now that we can write the power harvested by the receiver $R$ as follows:
\begin{eqnarray}
P({\cal C}(\mathbf{x}), R) & = & \gamma  \| \mathbf{E}({\cal C}(\mathbf{x}), R) \|^2 \nonumber \\
& = & \gamma  \left\| \sum_{C \in {\cal C}} \mathbf{x}_C  \mathbf{E}(C, R) \right\|^2 \nonumber \\
& = & \left( \sum_{C \in {\cal C}} \mathbf{x}_C  \sqrt{\gamma}  \mathbf{E}(C, R) \right)^T  \left( \sum_{C \in {\cal C}} \mathbf{x}_C  \sqrt{\gamma}  \mathbf{E}(C, R) \right) \nonumber \\
& = & (\mathbf{Q}^{(R)} \mathbf{x})^T \mathbf{Q}^{(R)} \mathbf{x}, \nonumber
\end{eqnarray}
where $(\cdot)^T$ denotes the transpose of a matrix or vector. Therefore, setting $\mathbf{H} \stackrel{def}{=} \sum_{R \in {\cal R}} \left(\mathbf{Q}^{(R)} \right)^T  \mathbf{Q}^{(R)}$, the solution to \texttt{MAX-POWER} is given by 
\begin{equation}
\mathbf{x}^* \in \arg\max_{\mathbf{x} \in [0, 1]^{m}} \mathbf{x}^T \mathbf{H} \mathbf{x}.
\end{equation}
It is worth noting that, in general, the maximization of a quadratic form is a non-convex quadratic program (even when $\mathbf{H}$ is positive semi-definite, which is the case here), hence cannot be solved in polynomial time. Nevertheless, by taking into account several properties and the special form of our problem, we are able to provide an efficient algorithm for \texttt{MAX-POWER}.

We first need the following elementary lemma that considerably reduces the size of the search space.

\begin{lemma} \label{lemma:0-1configuration}
If $\mathbf{x}^*$ is an optimal solution to \texttt{MAX-POWER}, then $\mathbf{x}^* \in \{0, 1\}^m$. In particular, there exists an optimal solution to \texttt{MAX-POWER} in which each charger either operates at full capacity or not at all.
%There is a configuration in $\mathbf{x}^* \in \{0, 1\}^m$ that is an optimal solution to the \texttt{MAX-POWER} problem. In particular, there exists an optimal solution to \texttt{MAX-POWER} in which each charger either operates at full capacity or not at all.
\end{lemma}
\begin{proof} Let $\mathbf{y} \in [0, 1]^m$ be such that \\ (a) $\mathbf{y} \in \arg\max_{\mathbf{x} \in [0, 1]^m} \mathbf{x}^T \mathbf{H} \mathbf{x}$ (i.e. it is an optimal solution to \texttt{MAX-POWER}) and (b) $\mathbf{y}$ has the minimum number of elements that are neither 0 nor 1; let also $k$ be the number of such elements, i.e. $k \stackrel{def}{=} |\{i: \mathbf{y}_i \notin \{0, 1\}\}|$. If $k=0$, then there is nothing to prove, so we will assume for the sake of contradiction that $k > 0$. Let $j$ be an index such that $\mathbf{y}_j \notin \{0, 1\}$ and define vectors $\mathbf{y}^{(0)}$ and $\mathbf{y}^{(1)}$ that are equal to $\mathbf{y}$ everywhere, except on position $j$, where $\mathbf{y}^{(0)}_j = 0$ and $\mathbf{y}^{(1)}_j = 1$. Notice now that, by the optimality and minimality assumptions on $\mathbf{y}$, we have that $\mathbf{y}^T \mathbf{H} \mathbf{y} > \left(\mathbf{y}^{(0)}\right)^T \mathbf{H} \mathbf{y}^{(0)}$ and $\mathbf{y}^T \mathbf{H} \mathbf{y} > \left(\mathbf{y}^{(1)}\right)^T \mathbf{H} \mathbf{y}^{(1)}$. 

On the other hand, for $z \in (0, 1)$, define the vector $\mathbf{y}^{(z)} \stackrel{def}{=} (1-z) \mathbf{y}^{(0)} + z \mathbf{y}^{(1)}$. In particular, $\mathbf{y}^{(z)}$ is the $m$-dimensional vector that is equal to $\mathbf{y}$ everywhere, except on position $j$, where $\mathbf{y}^{(z)}_j = z$. Consider now the function $f(z) = \left(\mathbf{y}^{(z)}\right)^T \mathbf{H} \mathbf{y}^{(z)}$. Notice that $f(z)$ is a single variable polynomial of degree 2, and it is a simple matter of calculus to show that its second derivative satisfies $\frac{d^2 f}{dz^2} = \mathbf{H}_{j, j} = \sum_{R \in {\cal R}} \gamma \|\mathbf{E}(C_j, R)\|^2 = P(C_j, {\cal R})$, which is strictly positive. But this implies that $f(z) < \max\{f(0), f(1)\}$, for all $z \in (0, 1)$. However, this is a contradiction, since we have already established that, by assumption, $f(\mathbf{y}_j) = \mathbf{y}^T \mathbf{H} \mathbf{y} > \max\{\left(\mathbf{y}^{(0)}\right)^T \mathbf{H} \mathbf{y}^{(0)}, \left(\mathbf{y}^{(1)}\right)^T \mathbf{H} \mathbf{y}^{(1)}\} = \max\{f(0), f(1)\}$. We conclude that in any optimal solution to \texttt{MAX-POWER} each charger either operates at full capacity or not at all. 
\qed
\end{proof}

We now prove a useful property of global maxima of the objective function $\mathbf{x}^T \mathbf{H} \mathbf{x}$ in $[0, 1]^m$. The proof uses properties of positive semi-definite (PSD) matrices (see \cite{993483} for an introduction to PSD matrices and their properties). We note here that, Lemma \ref{lemma:0-1configuration} and Theorem \ref{theorem:globaloptimum} below imply that any local maxima of the objective function $P({\cal C}(\mathbf{x}), R) = \left( \mathbf{Q} \mathbf{x}\right)^T \mathbf{Q} \mathbf{x}$ are also global maxima that belong to $\{0, 1\}^m$. In particular, this means that the gradient descent method  can be used to find a global maximum (i.e. an optimal solution to \texttt{MAX-POWER}). Nevertheless, in our evaluation, we used a pseudopolynomial distributed algorithm for computing the \emph{exact} optimum configuration for \texttt{MAX-POWER}, which is quite fast in practice. We present this algorithm later in this section.

\begin{theorem} \label{theorem:globaloptimum}
A configuration $\mathbf{x}^* \in \{0, 1\}^m$ is an optimal solution to \texttt{MAX-POWER} if and only if $P({\cal C}(\mathbf{x}^*), {\cal R}) \geq P({\cal C}(\mathbf{y}), {\cal R})$, for each $\mathbf{y}$ that comes from $\mathbf{x}$ by setting exactly one of its coordinates to either 0 or 1. 
\end{theorem}
\begin{proof} For a configuration $\mathbf{x} \in \{0, 1\}^m$ and for all $j \in [m]$ and $a \in \{0, 1\}$, define $\mathbf{x}^{(j, a)} \stackrel{def}{=} \mathbf{x} + (a-\mathbf{x}_j) \mathbf{e}_j$, where $\mathbf{e}_j$ is the $j$-th vector in the standard basis of $\mathbf{R}^m$. Notice that $\mathbf{x}^{(j, 1)}$ (respectively $\mathbf{x}^{(j, 0)}$) is the configuration that is identical to $\mathbf{x}$, with the only difference that charger $j$ operates at full capacity (respectively does not operate).

Clearly, if $\mathbf{x}^*$ is an optimal solution, then $P({\cal C}(\mathbf{x}^*), {\cal R}) \geq P({\cal C}(\mathbf{x}^{*(j, a)}), {\cal R})$, for any $j \in [m]$ and $a \in \{0, 1\}$. Therefore, it remains to prove the ``only if'' part of the Theorem. To this end, let $\mathbf{x}^*$ be such that $P({\cal C}(\mathbf{x}^*), {\cal R}) \geq P({\cal C}(\mathbf{x}^{*(j, a)}), {\cal R})$, for any $j \in [m]$ and $a \in \{0, 1\}$, and assume for the sake of contradiction that there is a configuration $\mathbf{z}$ such that $P({\cal C}(\mathbf{x}^*), {\cal R}) < P({\cal C}(\mathbf{z}), {\cal R})$.

By Lemma \ref{lemma:0-1configuration}, we only need to consider configurations in $\{0, 1\}^m$. Therefore, assume that $\mathbf{x}^* \in \{0, 1\}^m$ and $\mathbf{z} = \mathbf{x}^* + \sum_{j=1}^m a_j \mathbf{e}_j = \mathbf{x}^* + \mathbf{a}$, for some $\mathbf{a} \in \{-1, 0, 1\}^m$, such that $\mathbf{z}^T \mathbf{H} \mathbf{z} = P({\cal C}(\mathbf{z}), {\cal R}) > P({\cal C}(\mathbf{x}), {\cal R}) = \mathbf{x}^{*T} \mathbf{H} \mathbf{x}^*$. 

Note that, since $\mathbf{H}$ is symmetric, for any $j \in [m]$ we have $\mathbf{e}_j^T \mathbf{H} \mathbf{x}^* = \mathbf{x}^{*T} \mathbf{H} \mathbf{e}_j$, and so
\begin{equation} \label{eq:x_j}
(\mathbf{x}^* + a_j\mathbf{e}_j)^T \mathbf{H} (\mathbf{x}^* + a_j\mathbf{e}_j) = \mathbf{x}^{*T} \mathbf{H} \mathbf{x}^* + 2 a_j \mathbf{e}_j^T \mathbf{H} \mathbf{x}^* + a_j^2 \mathbf{H}_{j, j}.
\end{equation}
Rearranging, and using the assumption that $P({\cal C}(\mathbf{x}^*), {\cal R}) \geq P({\cal C}(\mathbf{x}^* + a_j\mathbf{e}_j), {\cal R})$, we get
\begin{equation} \label{ineq:j}
2 a_j \mathbf{e}_j^T \mathbf{H} \mathbf{x}^* + a_j^2 \mathbf{H}_{j, j} \leq 0,
\end{equation}
for any $j \in [m]$. By a similar computation, we have
\begin{eqnarray} 
\mathbf{z}^T \mathbf{H} \mathbf{z} & = & \mathbf{x}^{*T} \mathbf{H} \mathbf{x}^* + 2 \mathbf{a}^T \mathbf{H} \mathbf{x}^* + \mathbf{a}^T \mathbf{H} \mathbf{a} \nonumber \\
& = & \mathbf{x}^{*T} \mathbf{H} \mathbf{x}^* + \sum_{j=1}^m 2 a_j \mathbf{e}_j^T \mathbf{H} \mathbf{x} + \mathbf{a}^T \mathbf{H} \mathbf{a}. \label{eq:z}
\end{eqnarray}
Summing (\ref{ineq:j}) over all $j \in [m]$ and substituting in (\ref{eq:z}), we get 
\begin{equation} \label{ineq:diff}
\mathbf{z}^T \mathbf{H} \mathbf{z} - \mathbf{x}^{*T} \mathbf{H} \mathbf{x}^* \leq -\sum_{j=1}^m a_{j}^2 \mathbf{H}_{j, j} + \mathbf{a}^T \mathbf{H} \mathbf{a}.
\end{equation}
It is now a simple matter of algebra to show that the right hand side of the above inequality is non-positive. Indeed, let $\mathbf{H}'$ be the $m' \times m'$ principal submatrix of $\mathbf{H}$ corresponding to rows (and columns) $j$ for which $a_j \neq 0$ (in particular, $m'$ is the number of non-zero elements of $\mathbf{a}$). Clearly, since $\mathbf{H}$ is positive semi-definite, then so is $\mathbf{H}'$. Let $\lambda'_1 \geq \cdots \geq \lambda'_{m'} \geq 0$ and $\mathbf{v}'_1, \ldots, \mathbf{v}'_{m'}$ be the eigenvalues and eigenvectors of $\mathbf{H}'$. Now notice that $\sum_{j=1}^m a_{j}^2 \mathbf{H}_{j, j} = \sum_{j=1}^{m'} a_{j}^2 \mathbf{H}'_{j, j} = \mathrm{tr}(\mathbf{H}') = \sum_{j=1}^{m'} \lambda'_j$, where $\mathrm{tr}(\mathbf{H}')$ is the trace of $\mathbf{H}'$ and we have used the fact that the trace of a matrix is equal to the sum of its eigenvalues. Finally, since $\mathbf{a}$ is an orthonormal rotation of the vector $\sum_{j=1}^{m'} \mathbf{v}'_j$, we have $\mathbf{a}^T \mathbf{H} \mathbf{a} = \sum_{j=1}^{m'} \lambda'_j$. 

In view of the above, by inequality (\ref{ineq:diff}), we get that $\mathbf{z}^T \mathbf{H} \mathbf{z} - \mathbf{x}^{*T} \mathbf{H} \mathbf{x}^* \leq 0$, which is a contradiction. Therefore, we conclude that if $P({\cal C}(\mathbf{x}^*), {\cal R}) \geq P({\cal C}(\mathbf{x}^{*(j, a)}), {\cal R})$, for any $j \in [m]$ and $a \in \{0, 1\}$, then $\mathbf{x}^*$ is an optimal solution. 
\qed
\end{proof}

Lemma \ref{lemma:0-1configuration} and Theorem \ref{theorem:globaloptimum} suggest that the following distributed algorithm (which we call \texttt{IterativeMaxPower}) can be used to find an \emph{exact} optimum configuration for \texttt{MAX-POWER}: Initially, we begin from an arbitrary configuration in $\{0, 1\}^m$. In each subsequent step, we parse the set of chargers in order to find a charger $C \in {\cal C}$ such that the total power received by $R$ can be increased by flipping the operation level of $C$ (e.g. if $C$ operates at full capacity, it checks whether the received power is increased if it is not operational). The algorithm terminates if there is no such charger $C$. 

For a given placement of a family ${\cal C}$ of chargers and a family ${\cal R}$ of receivers, define $\delta({\cal C}, {\cal R}) \stackrel{def}{=} \min\{|P({\cal C}(\mathbf{x}), {\cal R}) - P({\cal C}(\mathbf{x}^{j, a}), {\cal R})|: \mathbf{x} \in \{0, 1\}^m, a \in \{0, 1\}, j \in [m] \}$. In particular, $\delta({\cal C}, {\cal R})$ is the minimal increment in the total received power that can be incurred by a single iteration of \texttt{IterativeMaxPower}. In addition, notice that every such iteration takes $O(m^3)$ time. Finally, given that the chargers and receivers satisfy the placement constraints of Subsection \ref{sec:assumptions}, a crude upper bound for the maximum total power is $n m^2 \gamma \beta^2 \frac{4 \pi^2}{\lambda^2} = O(n m^2)$. Therefore, we have the following:

\begin{theorem}
Given a family ${\cal C}$ of $m$ chargers and a family ${\cal R}$ of $n$ receivers that satisfy the placement constraints of Subsection \ref{sec:assumptions}, Algorithm \texttt{IterativeMaxPower} finds an optimal solution of \texttt{MAX-POWER} in $O\left( \frac{1}{\delta({\cal C}, {\cal R})} nm^5\right)$.
\end{theorem}

%%%%%%%%%%%%%%%%%%%%%%%%%%%%%%%%%%%%%%%%%%%%%%%%%%%%%%%%%%%%%%%%%%%%%%%%%%%%%%%%%%%%
%                     return optimal configuration
%%%%%%%%%%%%%%%%%%%%%%%%%%%%%%%%%%%%%%%%%%%%%%%%%%%%%%%%%%%%%%%%%%%%%%%%%%%%%%%%%%%%
\begin{algorithm}[h]
	%\DontPrintSemicolon
	\SetKwInOut{Input}{Input}\SetKwInOut{Output}{Output}
	%\SetKwComment{Comment}////\textit{Interaction between agents $i$ and $j$.}\\
	%\vspace{4pt}
	\Input{$dist, {\cal{R,\ C}}$, $communication\_range$}
	\Output{ $\mathbf{x}$}
	\Begin{
	$\mathbf{x}\in \{0,1\}^m$ is a random initial charger configuration\; 
	\While{$\exists C_j\in {\cal C}:P({\cal C}(\mathbf{x}), {\cal R}) < P({\cal C}(\mathbf{x}^{(j, a)}), {\cal R}),\ a\in\{0,1\}$}{
	choose randomly a charger $C_j\in {\cal C}$\;
	${\cal R}_{C_j}=0$ \;
	\ForEach{$R\in {\cal R}$  } {
	\If{$dist(C_j,R)\leq communication\_range$}{
		${\cal R}_{C_j}={\cal R}_{C_j}\cup R$\quad
		\quad\quad\quad\quad\quad\quad\quad\quad\quad\quad\quad\quad
		\quad\quad\quad\quad\quad\quad\quad\quad\quad
		\quad\quad\quad\quad\quad\quad //at this point $C_j$ communicates with $R$ and receives $\mathbf{E}({\cal C}(\mathbf{x}), R)$\;
		
		}
	}	
	%compute $P({\cal C}(\mathbf{x}^{(j, a)}), {\cal R}^{local}_{C_j}),\ \forall a\in\{0,1\}$\;
	%compute	$P({\cal C}(\mathbf{x}^{(j, 1)}), {\cal R}_{local})$\;
	$\mathbf{x}_{C_j}= \arg\max_{a \in \{0, 1\}} P({\cal C}(\mathbf{x}^{(j, a)}), {\cal R}_{C_j})$\;
	}	
	\textbf{return} $\mathbf{x}$\;
	}
\caption{\texttt{IterativeMaxPower}}
\label{distributed}
\end{algorithm}

\noindent\emph{Note:} For the performance evaluation, we implemented \texttt{IterativeMaxPower} using different levels of knowledge of the wireless system. In particular, we define the \emph{communication range} of a charger as the maximum radius of the disc area within which it can send and receive \emph{messages} from nodes. Hence, a transmitter ignores any node that is outside its communication range. Whenever a charger $C_j$ communicates with a node $R$, the latter sends to $C_j$ the energy field vector $\mathbf{E}({\cal C}(\mathbf{x}), R)$, where $\mathbf{x}$ is the configuration at the time when the communication took place. This information is enough for the charger to compute $P({\cal C}(\mathbf{x}^{(j, a)}), R)$, for each $a \in \{0, 1\}$, since $P({\cal C}(\mathbf{x}^{(j, a)}), R) = \gamma \|\mathbf{E}({\cal C}(\mathbf{x}), R) + (a - \mathbf{x}_j) \mathbf{E}(C_j, R) \|^2$. By using the above it is easy to compute $P({\cal C}(\mathbf{x}^{(j, a)}), {\cal R}_{C_j})$ for each $a \in \{0, 1\}$, where ${\cal R}_{C_j}\subseteq {\cal R}$ includes the nodes in the communication range of the charger $C_j$. The pseudocode of \texttt{IterativeMaxPower} can be found in Algorithm \ref{distributed} that simulates the distributed process in order to have an output. 

\texttt{IterativeMaxPower} is a distributed algorithm which can be applied in any Wireless Sensor Network (WSN) that supports WPT functionalities. The algorithm provides a mechanism that enhances a WPT network to be automatically adjusted to any non-scheduled or sudden changes like failure for a number of chargers or receivers. It can also be applied in dynamic networks and keep the quality of power charging in a very high level, while chargers and sensors are either added, removed or change their position. \texttt{IterativeMaxPower} achieves a better QoS in WPT with less power waste even though there is a trade-off between power consumption and the power that is harvested to sustain the network functional.

\section{Maximum $k$-Minimum Guarantee}
In this section, we present our algorithmic solutions to \texttt{MAX-$k$MIN-GUARANTEE}. This is more general than \texttt{MAX-POWER} and, even though we believe that it is computationally hard, we were unable to prove this formally. It is worth noting that the hardness of this problem does not lie in the computation of the minimum power among all $k$-set of receivers for a given configuration $\mathbf{x}$. It is not hard to see that this quantity is equal to the sum of powers of nodes having the $k$ minimum powers. 

\begin{figure}[h]
    \begin{subfigure}[b]{\columnwidth}
    \centering
        \includegraphics[width=0.9\textwidth]{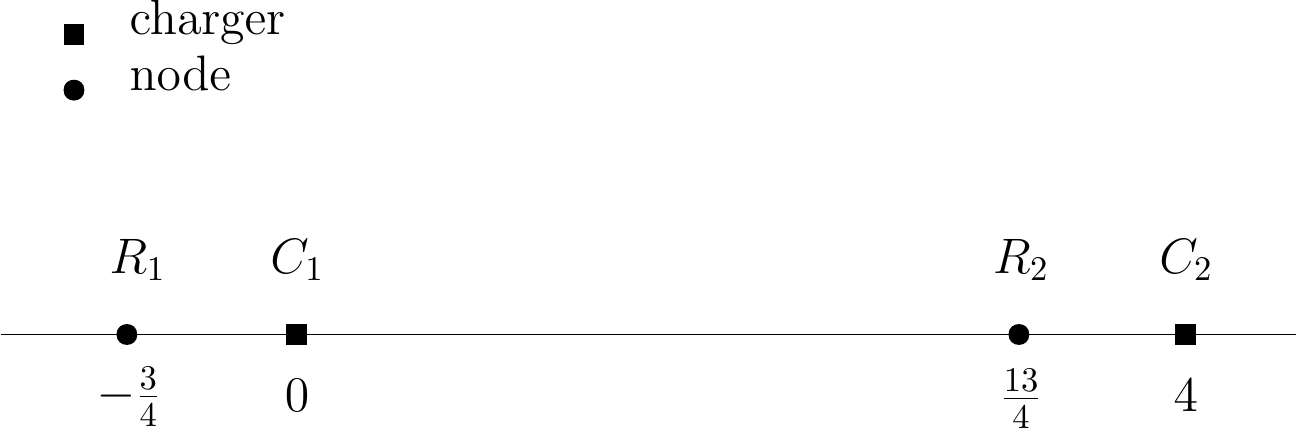}
        \captionsetup{justification=centering}
        \caption{Chargers' and nodes' placement on a straight line at points $(0, 0), (0, 4)$ and $(0,-\frac{3}{4}),(0,\frac{13}{4})$ respectively. }
        \label{fig:counterdeployment}
    \end{subfigure}
    \centering
    \begin{subfigure}[b]{\columnwidth}
        \includegraphics[width=\textwidth]{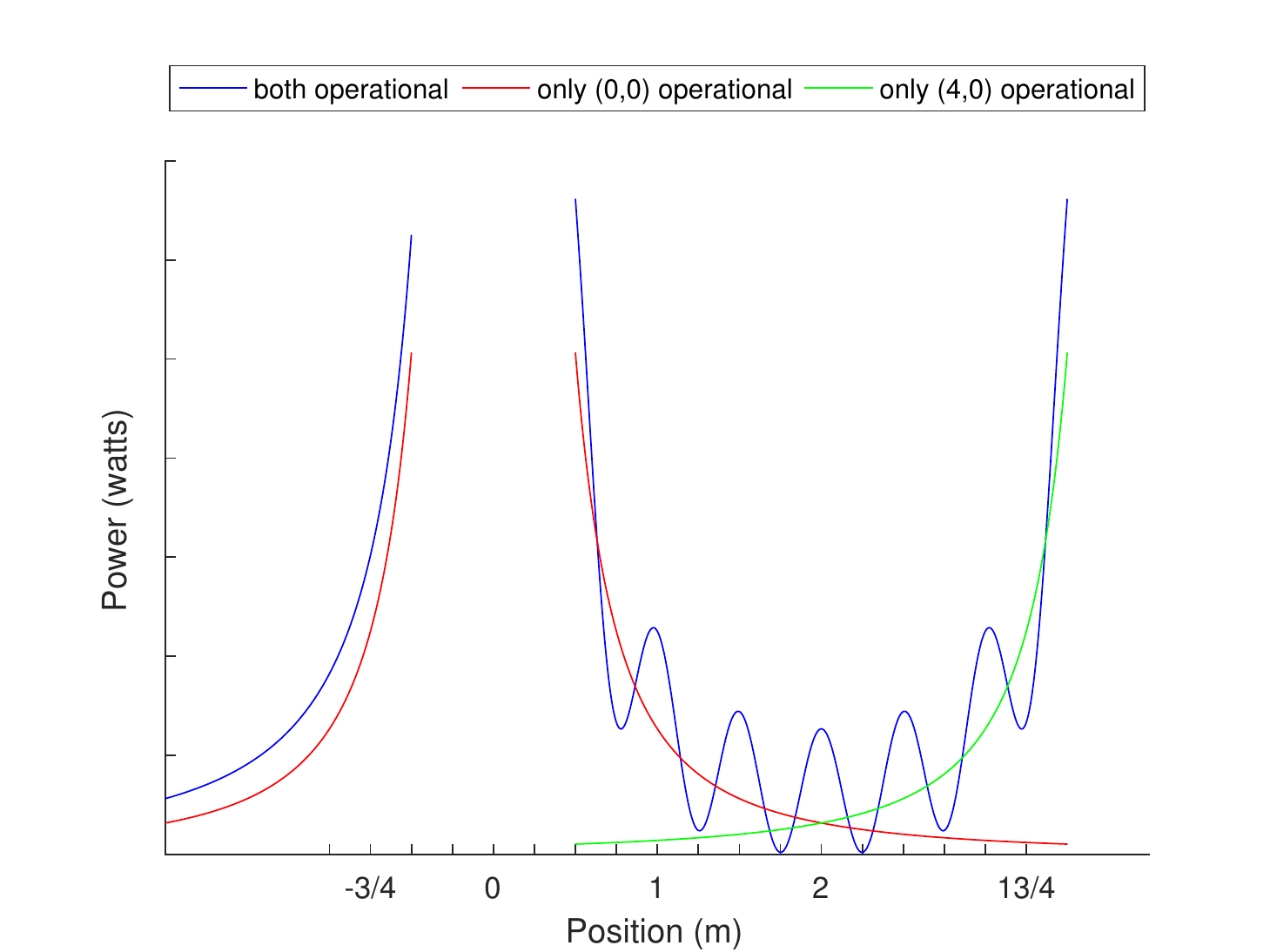}
        \caption{The power distribution in the straight line from the two chargers. Different curves represent different operation levels of the chargers.}
        \label{fig:counterexample}
    \end{subfigure}
    ~ %add desired spacing between images, e. g. ~, \quad, \qquad, \hfill etc. 
    %(or a blank line to force the subfigure onto a new line)
    \caption{Counter-example for fractional operation level of the chargers in \texttt{MAX-$k$MIN-GUARANTEE} problem. }\label{fig:figcounterexample}
\end{figure}

On the other hand, exhaustive search algorithms do not work in the general case. In particular, Lemma \ref{lemma:0-1configuration} does not apply, as we show in the following example in which it is established that fractional operation power level of the transmitters can achieve better performance than $x\in \{0,1\}^m$. Assume again that there are two transmitters $C_1$ and $C_2$ placed at points $(0, 0)$ and $(4, 0)$ and two receivers $R_1$ at $(-\frac{3}{4}, 0)$ and $R_2$ at $(\frac{13}{4}, 0)$ respectively (see Fig. \ref{fig:counterdeployment}). Also, assume that all constants in the above model are set to 1, i.e. $\lambda = \beta = \gamma =1$. Fig \ref{fig:counterexample} demonstrates the example on how the fractional operation level of $C_1$ can increase the $\min\{P(\{C_1,C_2\},\{R_1\}),P(\{C_1,C_2\},\{R_2\})\}$. So, if $x\in\{0,1\}^2$ there are three operation cases (obviously when both chargers are closed is excluded). Firstly, if only $C_1$ is on, then $P(C_1,R_1)=(\frac{4}{3})^2=1.77$ and $P(C_1,R_2)=(\frac{4}{13})^2=0.094$. On the other hand if $C_2$ is only on, the $\min\{P(C_2,R_1),P(C_2,R_2)\}=(\frac{1}{4})^2=0.0625$. In the final case, where $x_{C_1}=x_{C_2}=1$, the power at the points is $P(\{C_1,C_2\},R_1)=(\frac{4}{3}+\frac{4}{19})^2=2.38$ and $P(\{C_1,C_2\},R_2)=(\frac{4}{3}-\frac{4}{13})^2=1.025$. Clearly if the power from $R_1$ is reduced by a small amount $e$, then $P(\{C_1,C_2\},R_2)$ achieves a better minimum and at the same time $P(\{C_1,C_2\},R_1)$ remains high.

In view of the above hardness indications, we consider a relaxation of \texttt{MAX-$k$MIN-GUARANTEE} in which we are only interested in optimal configurations in $\{0, 1\}^m$, i.e. configurations in which each charger is either full operational, or does not operate.

%%%%%%%%%%%%%%%%%%%%%%%%%%%%%%%%%%%%%%%%%%%%%%%%%%%%%%%%%%%%%%%%%%%%%%%%%%%%%%%%%%%%
%                     OPtimal
%%%%%%%%%%%%%%%%%%%%%%%%%%%%%%%%%%%%%%%%%%%%%%%%%%%%%%%%%%%%%%%%%%%%%%%%%%%%%%%%%%%%
\begin{algorithm}[!]
	%\DontPrintSemicolon
	\SetKwInOut{Input}{Input}\SetKwInOut{Output}{Output}
	%\SetKwComment{Comment}////\textit{Interaction between agents $i$ and $j$.}\\
	%\vspace{4pt}
	\Input{$ dist, {\cal{R,\ C}},k$}
	\Output{$\mathbf{x}^{OPT}$}
	\Begin{
	$max\_power\_of\_k\_set=0$\;
	\ForEach{$ \mathbf{x}\in \{0,1\}^m$}
	{\ForEach{$  R\in {\cal{R}}$}
	{
	$\mathbf{p}(R)=P({\cal{C}}(\mathbf{x}),R)$\;}%=Power\ Received(x, dist, R, {\cal{C}})$\;}
	%$sort(P)$\;
	%$sum\_of\_k\_set=sum(P(1:k))$\;
	$sort(\mathbf{p},asc)$\;
	%$sum\_of\_k\_set=\sum (k\ min\ values\ of\ \mathbf{p})$\;
	\If{($\sum_{i=1}^k  \mathbf{p}(i) \geq max\_power\_of\_k\_set$)}
	{$max\_power\_of\_k\_set= \sum_{i=1}^k  \mathbf{p}(i)$\;
    $\mathbf{x}^{OPT}=\mathbf{x}$\;
	}
	}
	\textbf{return} $\mathbf{x}^{OPT}$\;
	}
\caption{Brute-force (OPT)}
\label{optimal}
\end{algorithm}

%%%%%%%%%%%%%%%%%%%%%%%%%%%%%%%%%%%%%%%%%%%%%%%%%%%%%%%%%%%%%%%%%%%%%%%%%%%%%%%%%%%%
%                     GREEDY
%%%%%%%%%%%%%%%%%%%%%%%%%%%%%%%%%%%%%%%%%%%%%%%%%%%%%%%%%%%%%%%%%%%%%%%%%%%%%%%%%%%%
\begin{algorithm}[h]
	%\DontPrintSemicolon
	\SetKwInOut{Input}{Input}\SetKwInOut{Output}{Output}
	%\SetKwComment{Comment}////\textit{Interaction between agents $i$ and $j$.}\\
	%\vspace{4pt}
	\Input{$dist, {\cal{R,\ C}},k$}
	\Output{Chargers configuration }
	\Begin{
	choose an arbitrary $\mathbf{x}\in \{0,1\}^m$\;
	$y=a\ random\ chargers\ sequence$\;
	\For{$i=1\rightarrow length(y)$}
	{\ForEach{$ R\in {\cal{R}}$}
	{
	%//open the charger\; 
	$\mathbf{p1}(R)=P({\cal{C}}(\mathbf{x}^{(y(i), 1)}),R)$\;%=Power\ Received(x, dist, r, {\cal{T}})$\;
	%//close the charger\; 
	$\mathbf{p0}(R)=P({\cal{C}}(\mathbf{x}^{(y(i), 0)}),R)$\;}%=Power\ Received(x, dist, r, {\cal{T}})$\;}
	$sort(\mathbf{p1},asc)$\;
	$sort(\mathbf{p0},asc)$\;
	%$sum\_k\_set_{1}=sum (k\ min\ values\ of\ P_{1})$\;
	%$sum\_k\_set_{0}=sum (k\ min\ values\ of\ P_{0})$\;
	%\eIf{($sum\_of\_k\_set_{0}\leq sum\_of\_k\_set_{1}$)}
	\eIf{($\sum_{j=1}^k  \mathbf{p0}(j) \leq \sum_{j=1}^k  \mathbf{p1}(j)$)}
	{%//we open the charger\;
    $\mathbf{x}_{y(i)}=1$\;
    %$max\_sum\_of\_k\_set=sum\_k\_set_{open}$\;
	}{%//we close the charger\;
    $\mathbf{x}_{y(i)}=0$\;
    %$max\_sum\_of\_k\_set=sum\_k\_set_{close}$\;
    }
	}\textbf{return} $\mathbf{x}$
	}
\caption{Greedy (GRE)}
\label{greedy}
\end{algorithm}

\textbf{Optimal configuration in $\{0, 1\}^m$.} We consider the following exhaustive search solution, which we use as a measure of comparison for our heuristics. In particular, \textit{Optimal Algorithm (OPT)} uses brute force to find an optimal solution. Due to its high time complexity $O(2^m)$, it is not practical when the problem size tends to grow. OPT can serve as a performance upper bound when benchmarking our algorithms.

Algorithm \ref{optimal} crushingly searches (without taking anything into consideration) among all the possible configurations of the chargers the one that maximizes the cumulative power of the $k-$set  of nodes with the least power received.

\textbf{Greedy Algorithm (GRE)}: The algorithm initiates from a random configuration which through an iterative process improves. The algorithm's decision for the chargers' operation level is the one that contributes more to the cumulative received power of the $k-$set that consist of the nodes with the least power received. Although the heuristic disambiguation can sometimes perform badly, it may yield locally optimal solution that approximates the global optimum in reasonable time. Clearly, time complexity depends on the number the algorithm checks to calibrate the chargers. 

%%%%%%%%%%%%%%%%%%%%%%%%%%%%%%%%%%%%%%%%%%%%%%%%%%%%%%%%%%%%%%%%%%%%%%%%%%%%%%%%%%%%
%                     SAMPLE
%%%%%%%%%%%%%%%%%%%%%%%%%%%%%%%%%%%%%%%%%%%%%%%%%%%%%%%%%%%%%%%%%%%%%%%%%%%%%%%%%%%%
\begin{algorithm}[h!]
	%\DontPrintSemicolon
	\SetKwInOut{Input}{Input}\SetKwInOut{Output}{Output}
	%\SetKwComment{Comment}////\textit{Interaction between agents $i$ and $j$.}\\
	%\vspace{4pt}
	\Input{$ dist, {\cal{R,\ C}},k$}
	\Output{$\mathbf{x}$}
	\Begin{
	choose randomly $\sigma$ $k$-sets of nodes $\{k_1, k_2, ..., k_\sigma\}$\;
	\For{$i=1 \rightarrow \sigma$}{$\mathbf{x^i}=\texttt{IterativeMaxPower}(dist, k_i, {\cal{ C}},open)$\;}
	$perm=a\ random\ permutation\ of\ chargers$\;
	\For{$j=1\rightarrow m$}
	{$gain\_0=0$\;
	$gain\_1=0$\;
	\For{$i=1\rightarrow \sigma$}
	{
	%//we close the charger\;
	$\mathbf{p0}(i)=P({\cal{C}}(\mathbf{x^i}^{((perm(j), 0)}),k_i)$\;%=Power\ Received(x_i, dist, k_i, {\cal{C}})$\;
	%//we open the charger\;
	$\mathbf{p1}(i)=P({\cal{C}}(\mathbf{x^i}^{((perm(j), 1)}),k_i)$\;%=Power\ Received(x_i, dist, k_i, {\cal{C}})$\;
	\eIf{($\mathbf{p1}(i) \leq \mathbf{p0}(i)$)}
	{$gain\_0=gain\_0+\mathbf{p0}(i) -\mathbf{p1}(i) $\;
	}{$gain\_1=gain\_1+\mathbf{p1}(i) -\mathbf{p0}(i) $\;
	}
	}
	\eIf{($gain\_1 \leq gain\_0$)}
	{$\forall \ i\in [\sigma],\mathbf{x}^i_{perm(j)}=0$\;
	}{$\forall \ i\in [\sigma],\mathbf{x}^i_{perm(j)}=1$\;
	}
	}
	\textbf{return} any $\mathbf{x}^i:i\in[\sigma]$
	}
\caption{Sampling (SAM)}
\label{sample}
\end{algorithm}

%%%%%%%%%%%%%%%%%%%%%%%%%%%%%%%%%%%%%%%%%%%%%%%%%%%%%%%%%%%%%%%%%%%%%%%%%%%%%%%%%%%%
%                     FUSION
%%%%%%%%%%%%%%%%%%%%%%%%%%%%%%%%%%%%%%%%%%%%%%%%%%%%%%%%%%%%%%%%%%%%%%%%%%%%%%%%%%%%
\begin{algorithm}[h!]
	%\DontPrintSemicolon
	\SetKwInOut{Input}{Input}\SetKwInOut{Output}{Output}
	%\SetKwComment{Comment}////\textit{Interaction between agents $i$ and $j$.}\\
	%\vspace{4pt}
	\Input{$ dist, {\cal{R,\ C}},k$}
	\Output{Chargers configuration }
	\Begin{
	//find optimal configuration for each node\;
	\ForEach{$  R\in {\cal{R}}$}{$\mathbf{x}^R=\texttt{IterativeMaxPower}(dist, R, {\cal{ C}},open)$\;}
	$perm=a\ random\ permutation\ of\ chargers$\;
	\For{$j=1\rightarrow m$}
	{\ForEach{$  R\in {\cal{R}}$}
	{
	%//open charger\;
	$\mathbf{p0}(R)=P({\cal{C}}(\mathbf{x}^{R((perm(j), 0)}),R)$\;%=Power\ Received(x_R, dist, R, {\cal{C}})$\;
	%//close charger\;
	$\mathbf{p1}(R)=P({\cal{C}}(\mathbf{x}^{R((perm(j), 1)}),R)$\;%=Power\ Received(x_R, dist, R, {\cal{C}})$\;
	}
	%$sort(P_{close})$\;
	%$sort(P_{open})$\;
	%$sum\_k\_set_{1}=sum (k\ min\ values\ of\ P_{1})$\;
	%$sum\_k\_set_{0}=sum (k\ min\ values\ of\ P_{0})$\;
	$sort(\mathbf{p1},asc)$\;
	$sort(\mathbf{p0},asc)$\;
	\eIf{($\sum_{i=1}^k  \mathbf{p1}(i) \leq \sum_{i=1}^k  \mathbf{p0}(i)$)}
	{$\forall R \in {\cal{R}},\ \mathbf{x}^R_{perm(j)}=0$\;
	}{$\forall R \in {\cal{R}},\ \mathbf{x}^R_{perm(j)}=1$\;
	}
	}
	\textbf{return} any $\mathbf{x}^R: R \in {\cal{R}}$
	}
\caption{Fusion (FUS)}
\label{fusion}
\end{algorithm}

Algorithm \ref{greedy} tries to find an optimal solution after a number of iterations. In each step, it randomly chooses a charger $y(i)$, computes the power received from each node $R$ and stores them in the vector $\mathbf{p1}$ when $y(i)$ is activated and at vector $\mathbf{p0}$ when is deactivated. Subsequently, GRE measures the cumulative received power of the $k$ nodes that receive the less power from $\mathbf{p1}$ and $\mathbf{p0}$ respectively. The operation level that is providing the larger amount of cumulative received power will be the algorithm's choice for the current step.

\textbf{Sampling Algorithm (SAM)}: Instead of going through all possible solutions extensively as the Brute-force algorithm does, and in order to avoid the greedy approach and its disadvantages, we propose a sampling heuristic. It randomly samples $\sigma\ k-$sets of nodes and aims to maximize the cumulative received power of all possible $k-$sets with the configuration that will come up from the sample. In this way, the algorithm overcomes the threat of a locally optimal solution but due to the random grouping, it doesn't take into account the nodes that form the $k-$set with the least received power.

Algorithm \ref{sample} consists of two phases. During the first phase it randomly chooses $\sigma\ k-$sets of nodes and finds their optimal chargers configuration $\mathbf{x}^i:i\in[1,\sigma]$ via Algorithm \ref{distributed}. The second phase is iterative. Within each iteration, it chooses a charger $perm(j)$ from a random permutation ($perm$) of the chargers set and measures the cumulative received power of each $k-$set that belongs to the sample when the charger is activated and when it is not. Then, it sums the gain when the charger is activated ($gain\_1$) and compares it with the gain when it is closed ($gain\_0$). The operation level with the larger sum of gain is decided as the final operation level of the charger. Thus, at the end of the process the $\sigma\ k-$sets share the same chargers configuration. Finally, the configuration of the sampled $k-$sets is SAM's output.

Two extensions of the above algorithm, which arise from the relaxation of the $\{0,1\}$ power configuration restriction, are the following:

\textit{Extension 1}: If we have the choice to configure the charger's transmitting power in $(0,1)$, then for the first extension we can set the operational level of the charger to be $\frac{gain\_1}{gain\_1+gain\_0}$ when $gain\_1 > gain\_0$ or $1-\frac{gain\_0}{gain\_1+gain\_0}$ when $gain\_0 > gain\_1$. This way, we configure the charger according to the beneficial units of power that it provides to the network. So, if we get $gain\_1=10$ and $gain\_0=5$ then the operational level of the charger is set to $66\%$.

\textit{Extension 2}: The criterion for the second extension is the number of samples that are benefited from that particular charger. Thus, we set the operational level of a charger at $\frac{\#\ of\ k-sets\ that\ the\ impact\ of\ the\ charger\ is\ positive}{\sigma}$. For example, if the charger's configuration is $1$ and it is beneficial for the $10$ out of $30$ samples, then its operational level would be $33\%$, in order to moderate the destructive interference of its transmitted power.  

\textbf{Fusion Algorithm (FUS)}: Our last algorithm tries to combine the advantages of the previous ones and at the same time to restrict their weaknesses. FUS initiates by having the best chargers configuration for each node individually. Step by step it changes the operation level of one charger at the time. This way, the algorithm fuses the $n$ different configurations to one with respect to the received power sum of the $k$ nodes with the least power. FUS takes into consideration the received power for each node separately and aims to keep it as high as possible.

Algorithm \ref{fusion} consists of two phases. During the first phase, it finds the optimal chargers configuration $\mathbf{x}^R$ via algorithm \ref{distributed} for each node $R$ individually. During the second phase, it proceeds iteratively. Within each iteration, it chooses a charger $perm(j)$ from a random permutation of the chargers set and measures the power of each node when it is activated and when its not. Those power values are stored in two vectors, $\mathbf{p1}$ and $\mathbf{p0}$ respectively. In the sequel, it sums the $k$ nodes with the least power for each operation level and compares them. The operation level of the charger with the larger sum will be chosen. That way FUS changes the charger's operation level to the one that has the less negative impact on the $k-$set with the least received power that has risen. At the end of the algorithm, all the nodes share the same chargers configuration.

\begin{table*}[t!]
\centering
\label{my-label}
\begin{tabular}{|l|l|l|l|l|}
\hline
\textbf{Algorithm} & \textbf{Assumption} & \textbf{Knowledge} & \begin{tabular}[c]{@{}l@{}}\textbf{Running} \\ \textbf{Time}\end{tabular} & \textbf{Performance} \\ \hline
\textbf{Greedy} & \begin{tabular}[c]{@{}l@{}}Initiate from a random \\ configuration for chargers\end{tabular} & \begin{tabular}[c]{@{}l@{}}checks all the nodes\\  of the network\end{tabular} & Fast & Low \\ \hline
\textbf{Sampling} & \begin{tabular}[c]{@{}l@{}}Initiate with $\sigma$ different\\  configuration, optimal for each sample\end{tabular} & \begin{tabular}[c]{@{}l@{}}checks only the nodes \\ of the sample\end{tabular} & Average & \begin{tabular}[c]{@{}l@{}}Low for small\\  $k$, Average\end{tabular} \\ \hline
\textbf{Fusion} & \begin{tabular}[c]{@{}l@{}}Initiate with n different \\ configuration, optimal for each node\end{tabular} & \begin{tabular}[c]{@{}l@{}}checks all the nodes\\  of the network\end{tabular} & Slow & High \\ \hline
\end{tabular}
\caption{Summary Table}
\end{table*}

\section{Performance Evaluation}\label{evaluation}

\begin{figure}
    \centering
    \begin{subfigure}[b]{0.9\columnwidth}
        \includegraphics[width=\textwidth]{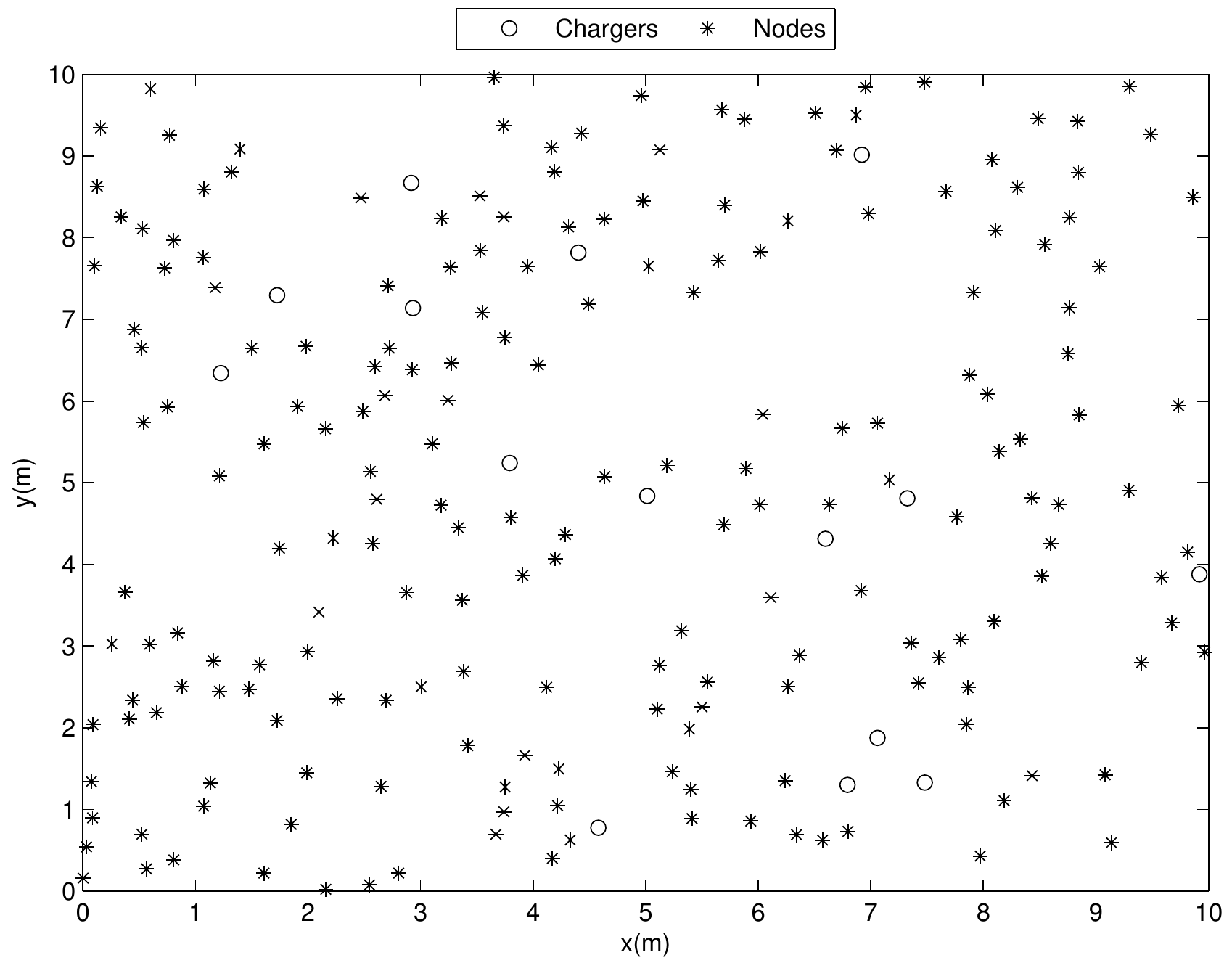}
        \caption{Chargers' and nodes' deployment.}
        \label{fig:deployment}
    \end{subfigure}
    ~ %add desired spacing between images, e. g. ~, \quad, \qquad, \hfill etc. 
      %(or a blank line to force the subfigure onto a new line)
%    \begin{subfigure}[b]{0.9\columnwidth}
%        \includegraphics[width=\textwidth]{quadratic.eps}
%        \caption{Cumulative power over time for different knowledge levels.}
%        \label{fig:quadratic}
%    \end{subfigure}
    ~ %add desired spacing between images, e. g. ~, \quad, \qquad, \hfill etc. 
    %(or a blank line to force the subfigure onto a new line)
    \begin{subfigure}[b]{0.9\columnwidth}
        \includegraphics[width=\textwidth]{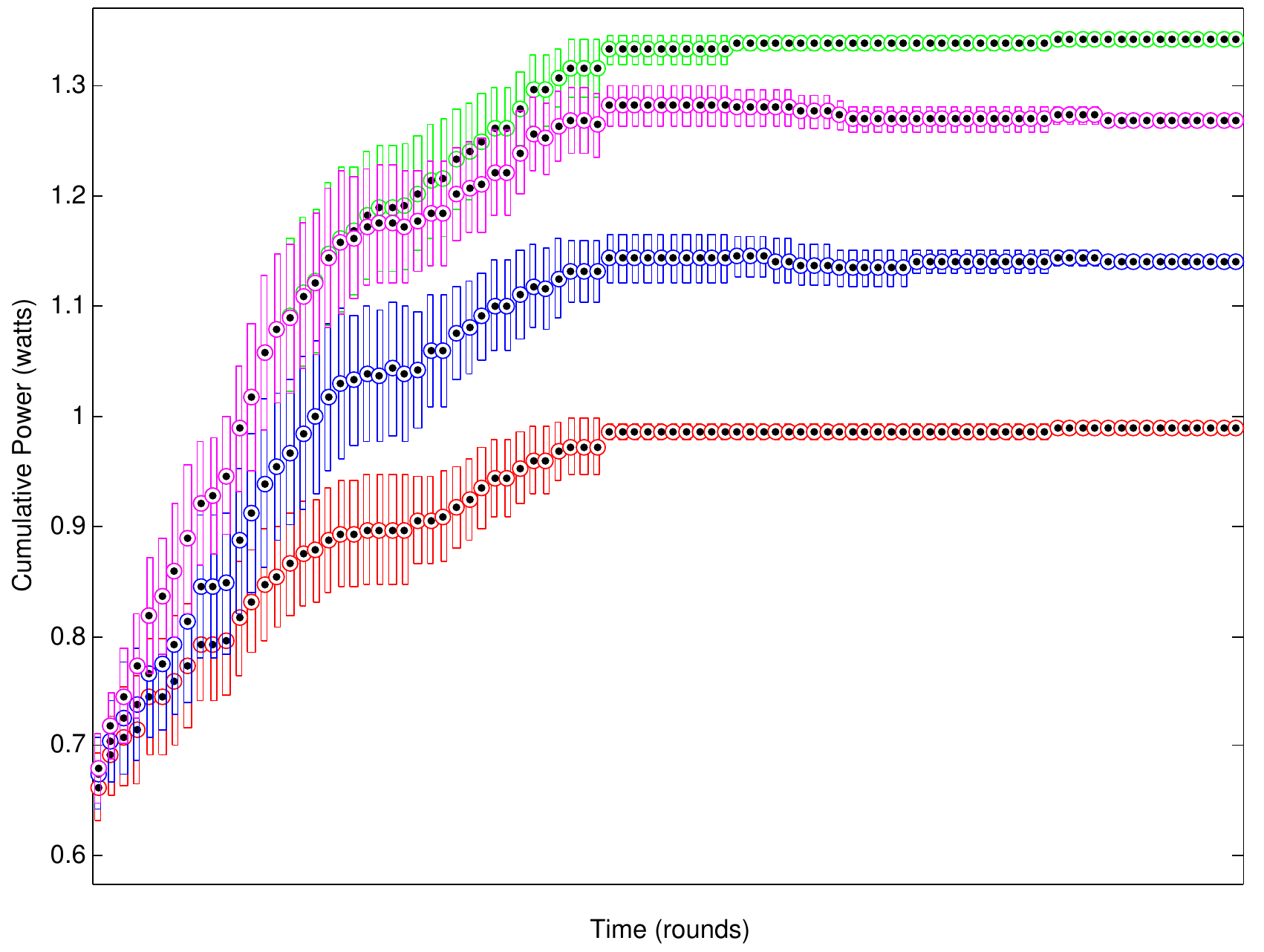}
        \captionsetup{justification=centering}
        \caption{Confidence intervals for the cumulative power over time for different knowledge levels.\\open \fcolorbox{black}{green}{\rule{0pt}{3pt}\rule{0pt}{0pt}}\quad 1 \fcolorbox{black}{magenta}{\rule{0pt}{3pt}\rule{0pt}{0pt}}\quad 0.6, \fcolorbox{black}{blue}{\rule{0pt}{3pt}\rule{0pt}{0pt}}\quad 0.4 \fcolorbox{black}{red}{\rule{0pt}{3pt}\rule{0pt}{0pt}}}
        \label{fig:ci}
    \end{subfigure}
    \caption{Deployment and cumulative power for \texttt{IterativeMaxPower}.}\label{fig:}
\end{figure}

We conducted simulations in order to evaluate our methods and provide useful insights for the WPT process, using Matlab R2016a. We consider a system that consists of chargers and nodes randomly deployed in a square field of 10m $\times$ 10m. Each charger's antenna delivers power equal to 2W and has 2 dbi gain, while the gain of the receiver's antenna is 1 dbi. The number of chargers and nodes is set to 15 and 200 respectively and all the experiments run on the same system where the wavelength is 29cm. Fig \ref{fig:deployment} depicts an instance of the deployment. For statistical smoothness, we conducted each simulation 100 times. Even though the statistical analysis of the findings demonstrates very high concentration around the mean, in the following simulation results we also depict the confidence intervals. Actually, we provide the confidence intervals for 20 repetitions, in order to demonstrate an earlier convergence.

In this section, we provide our simulation results on three performance metrics: (a) power efficiency, (b) cumulative power fuelled into the system, (c) communication overhead (number of messages that have been sent or received from nodes) and (d) power balance (the variation of power among nodes).

\textbf{Power Efficiency:}
We consider the ratio of the received power by the nodes to the power fuelled by the chargers. For this particular simulation the varying number of chargers and the 20 nodes were deployed in a square field of 3m $\times$ 3m.

\begin{figure}[h]
    \centering
    \begin{subfigure}[b]{0.9\columnwidth}
        \includegraphics[width=\textwidth]{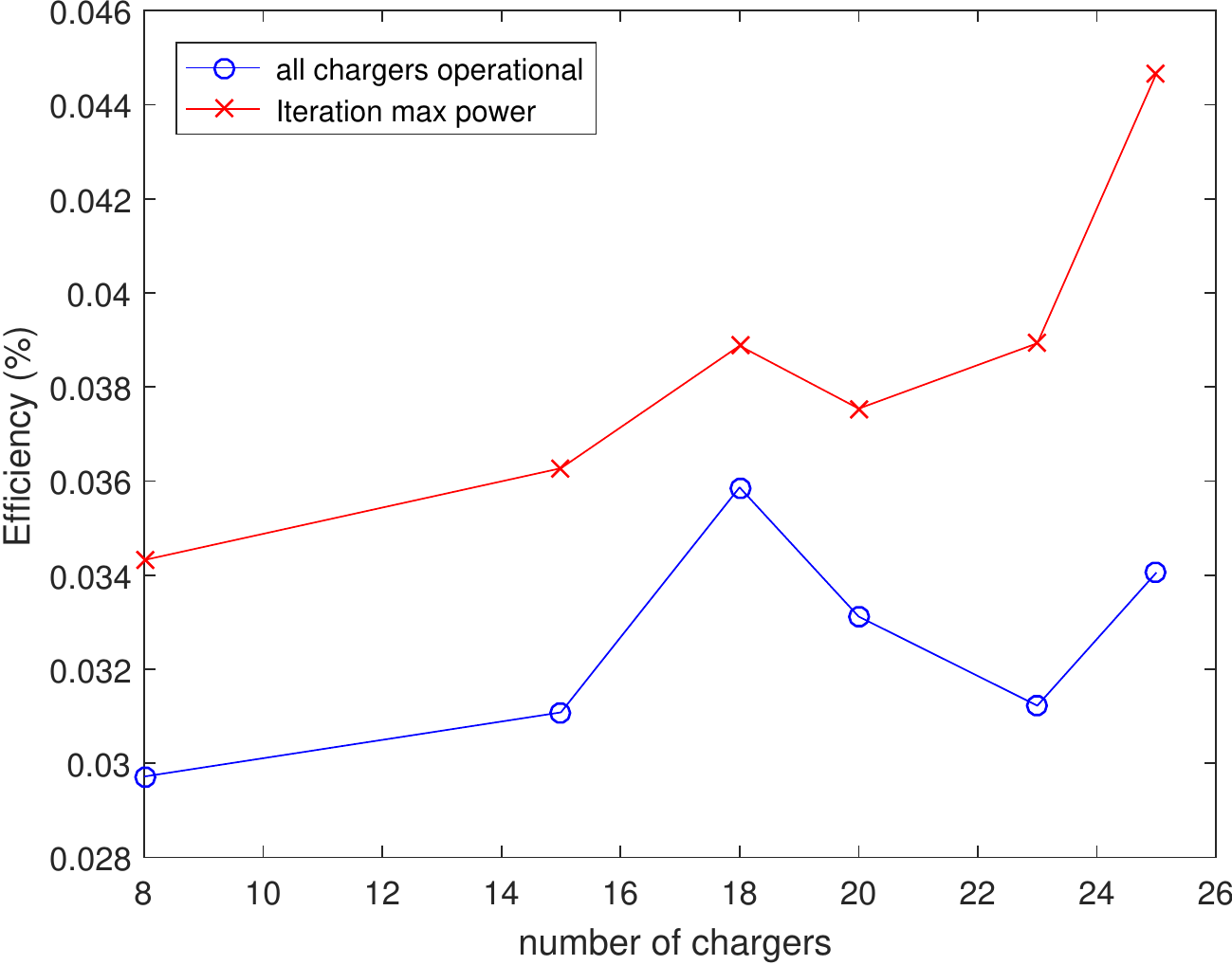}
    \end{subfigure}
    \caption{Power efficiency for different number of chargers.}\label{fig:efficiency}
\end{figure}

For the ``all chargers operational'' curve we set all the chargers operational (as scalar model suggests), while for the second one we apply \texttt{IterativeMaxPower}. The performance in terms of power efficiency of the two different approaches is depicted in Fig \ref{fig:efficiency}. We observe that the \texttt{IterativeMaxPower} achieves better performance as, not only does the configuration of the chargers provide more power to the network, but also saves energy from the turned off chargers. 

\textbf{Cumulative Power:}
In general, as the communication range of the chargers grows, the distributed algorithm \ref{distributed} achieves a near optimal solution. Fig \ref{fig:ci} depicts the cumulative received power by the nodes under varying the communication range of the chargers over time (for 90 rounds). In each round, a charger is chosen randomly and decides on its operation level. Fig \ref{fig:deployment} shows the corresponding deployment for both problems. 

As for the open communication range, the cumulative power never decreases over time like the others do. This is because every charger has global knowledge of the power exchange in the system. Indeed, when a charger has limited communication range, then its choice serves the nodes in the communication range that covers, but for the rest of the nodes, the result might be negative. On the other hand, at the first steps of the distributed algorithm, choices made from chargers with short communication range can benefit temporarily the cumulative power for the next steps, but in the end, the one with the global knowledge will perform better.

\begin{figure}[!]
    \centering
    \begin{subfigure}[b]{0.9\columnwidth}
        \includegraphics[width=\textwidth]{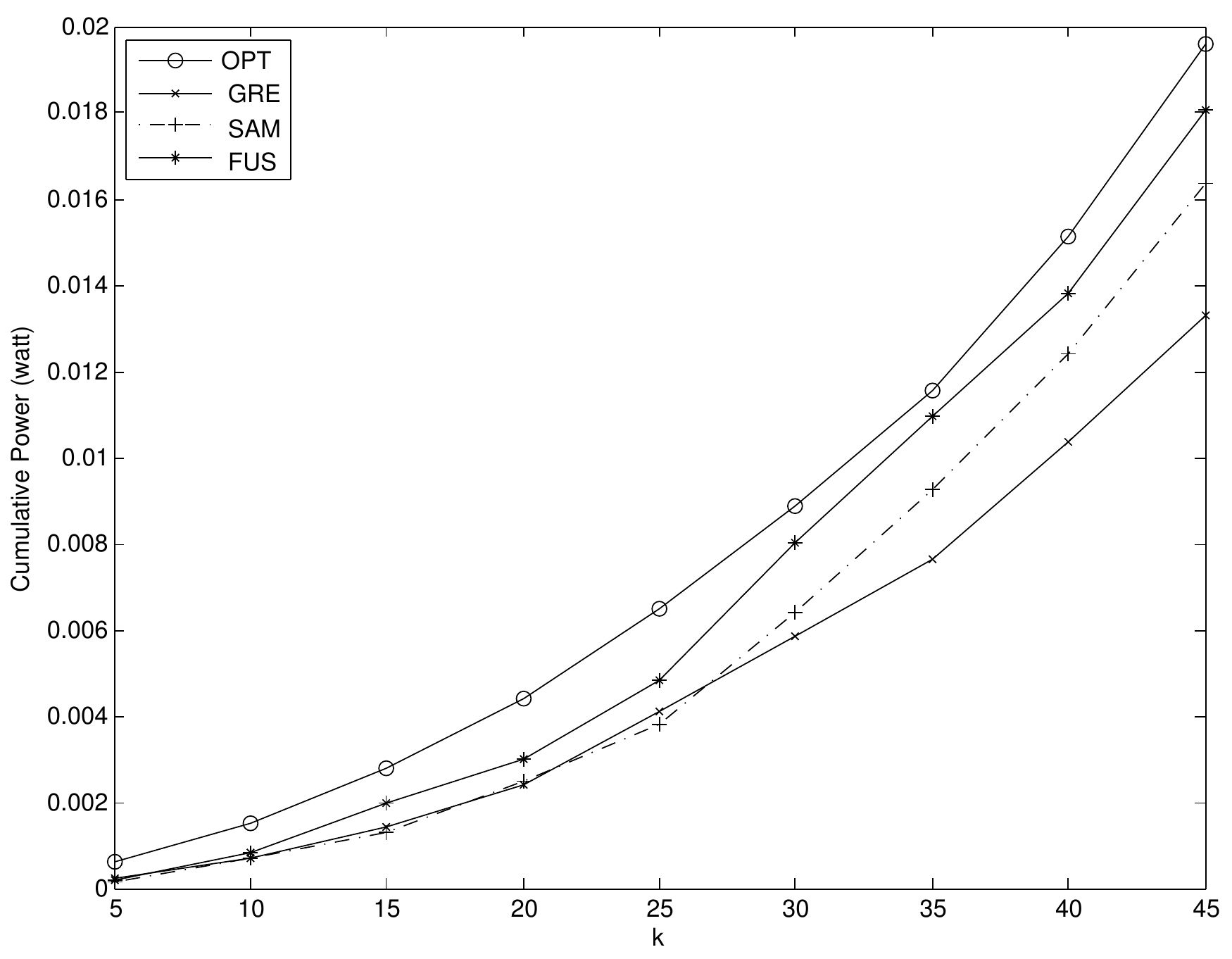}
        \caption{Cumulative power (of $k-$set with minimum power) for different values of $k$.}
        \label{fig:generalcp}
    \end{subfigure}
    ~ %add desired spacing between images, e. g. ~, \quad, \qquad, \hfill etc. 
    %(or a blank line to force the subfigure onto a new line)
    \begin{subfigure}[b]{0.9\columnwidth}
        \includegraphics[width=\textwidth]{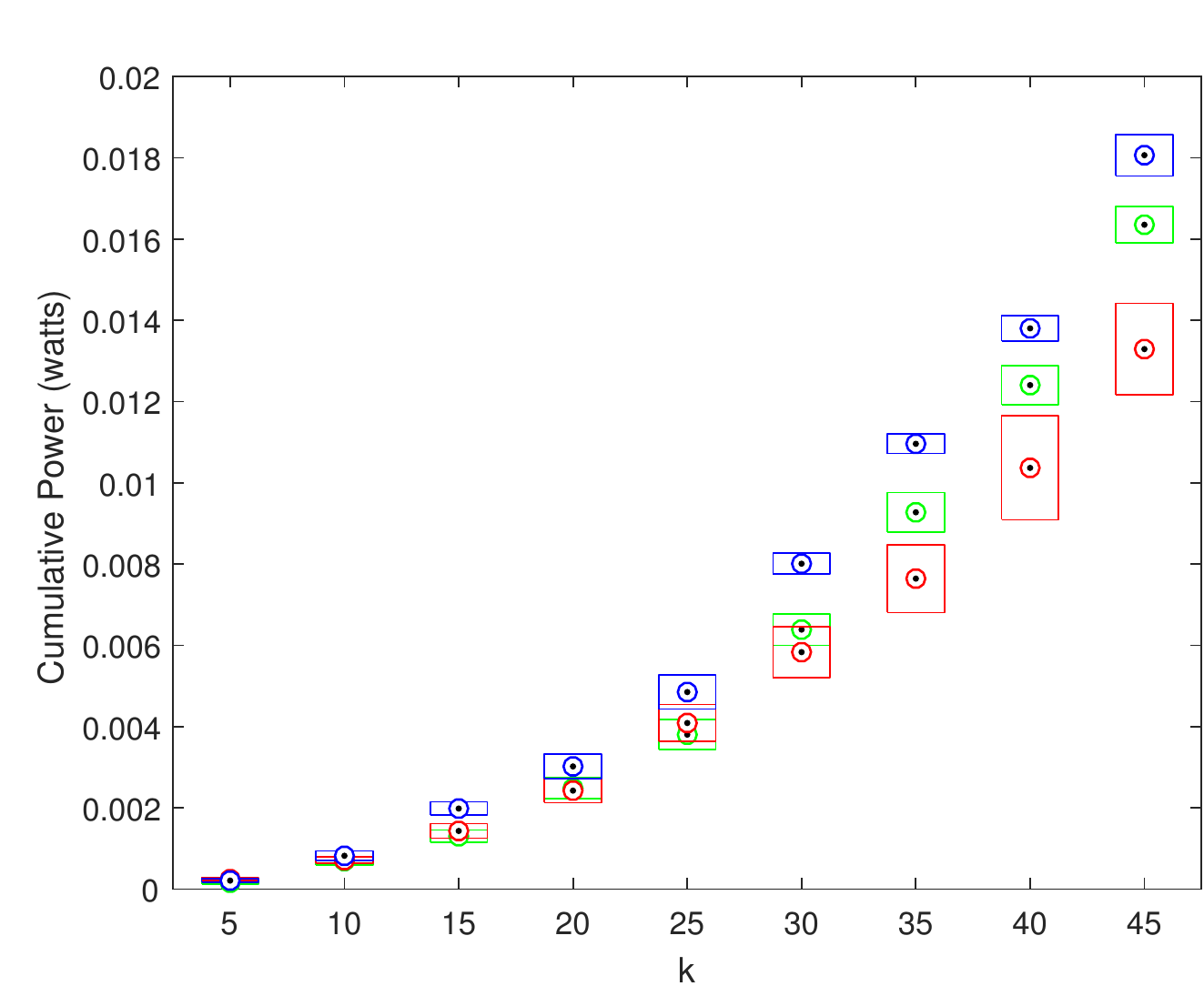}
        \captionsetup{justification=centering}
        \caption{Confidence intervals for \ref{fig:generalcp}.\\ GRE \fcolorbox{black}{red}{\rule{0pt}{3pt}\rule{0pt}{0pt}}\quad SAM \fcolorbox{black}{green}{\rule{0pt}{3pt}\rule{0pt}{0pt}}\quad FUS \fcolorbox{black}{blue}{\rule{0pt}{3pt}\rule{0pt}{0pt}}\quad }
        \label{fig:generalci}
    \end{subfigure}
    \begin{subfigure}[b]{0.9\columnwidth}
        \includegraphics[width=\textwidth]{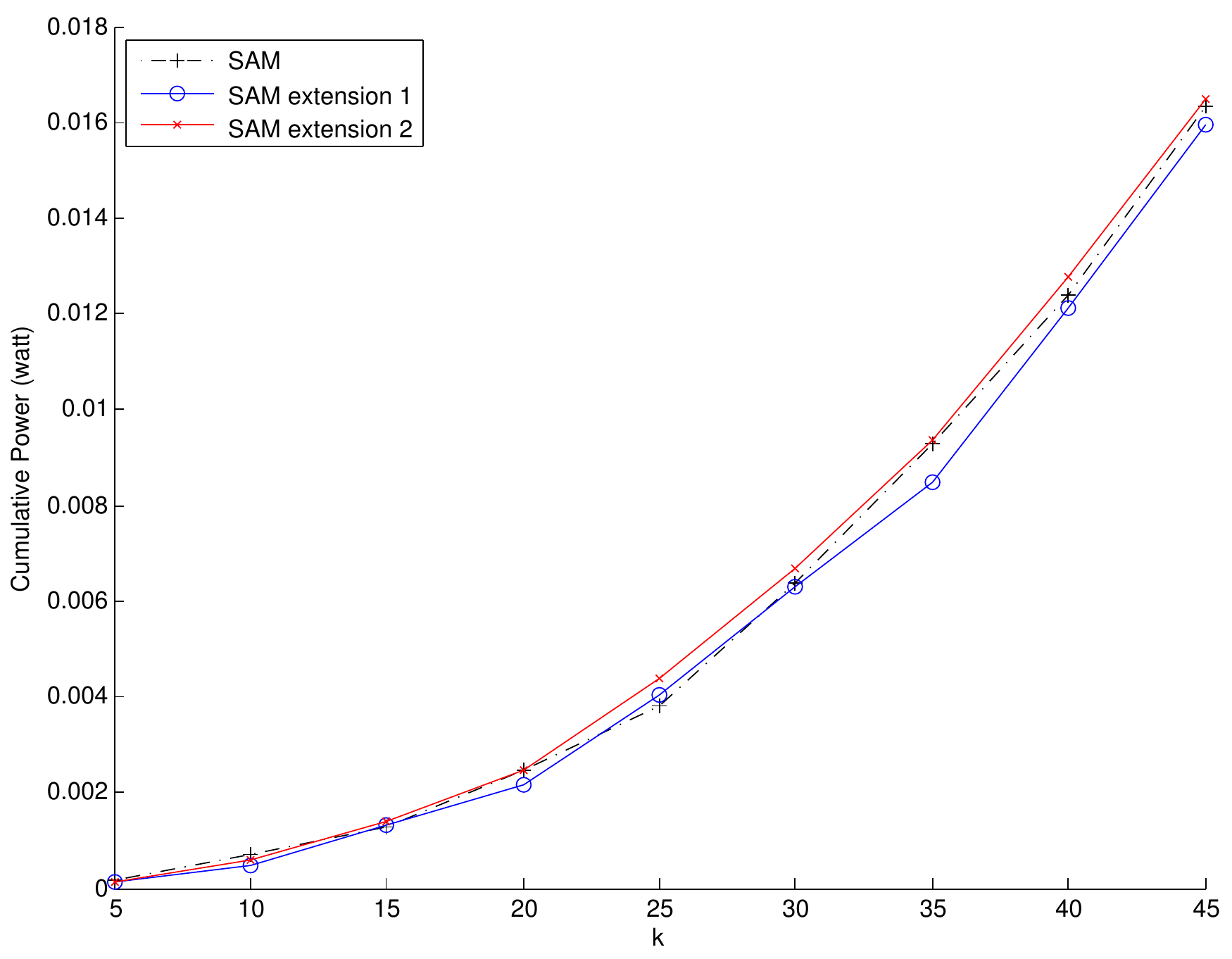}
        \captionsetup{justification=centering}
        \caption{Performance of SAM and its two extensions.}
        \label{fig:extensions}
    \end{subfigure}
    \caption{Performance of different algorithms.}\label{fig:general}
\end{figure}

%\fcolorbox{black}{magenta}{\rule{0pt}{6pt}\rule{6pt}{0pt}}\quad Text text text text text.

%\begin{figure}[!ht]
%  \caption{\fcolorbox{black}{magenta}{\rule{0pt}{6pt}\rule{6pt}{0pt}}\quad Text text text text text.}
%  \centering
    %\includegraphics[width=0.5\textwidth]{gull}
%\end{figure}

We also observe that when the communication range is 1m we achieve a near optimal solution. The reason behind this, is that the received power of the nodes far from the charger is reduced due to distance and they don't contribute that much to the cumulative received power. The confidence intervals which are also presented in Fig \ref{fig:ci} show a high concentration around the mean and the absence of overlaps for the last steps. 

For \texttt{MAX-$k$MIN-GUARANTEE} we are interested in the impact of the number $k$. Fig \ref{fig:generalcp} depicts the cumulative power of the $k$ nodes with the least power over various $k$. An interesting observation is that GRE performs better than SAM for low $k$. On the other hand for $k>27$ (as the population grows) we have a considerable improvement for SAM. In our simulations $\sigma=30$. Different values of $\sigma$ didn't provide any significant improvement, but if it is too small, then deal with a poor sample. As expected, FUS outperforms the other two algorithms and it is the one that approximates OPT better. This is because FUS checks the power received for all the nodes and has more than one initial chargers' configuration unlike GRE. The confidence intervals are also presented in the Fig \ref{fig:generalci}. As we claimed, the algorithms present high concentration around the mean and there are no overlaps as the value of $k$ grows.

In \ref{fig:extensions} we present the performance of the two heuristics that came up from the extension of SAM to a fractional operational level configuration of the chargers. The purpose of this simulation is to clarify any hidden aspects of a configuration that applies percentage operation for the chargers. Since the two algorithms have similar results to SAM we conclude that a different approach is needed in order to take advantage of the example \ref{fig:counterexample} outcome.

\begin{figure}[!]
    \centering
    ~ %add desired spacing between images, e. g. ~, \quad, \qquad, \hfill etc. 
      %(or a blank line to force the subfigure onto a new line)
    \begin{subfigure}[b]{0.9\columnwidth}
        \includegraphics[width=\textwidth]{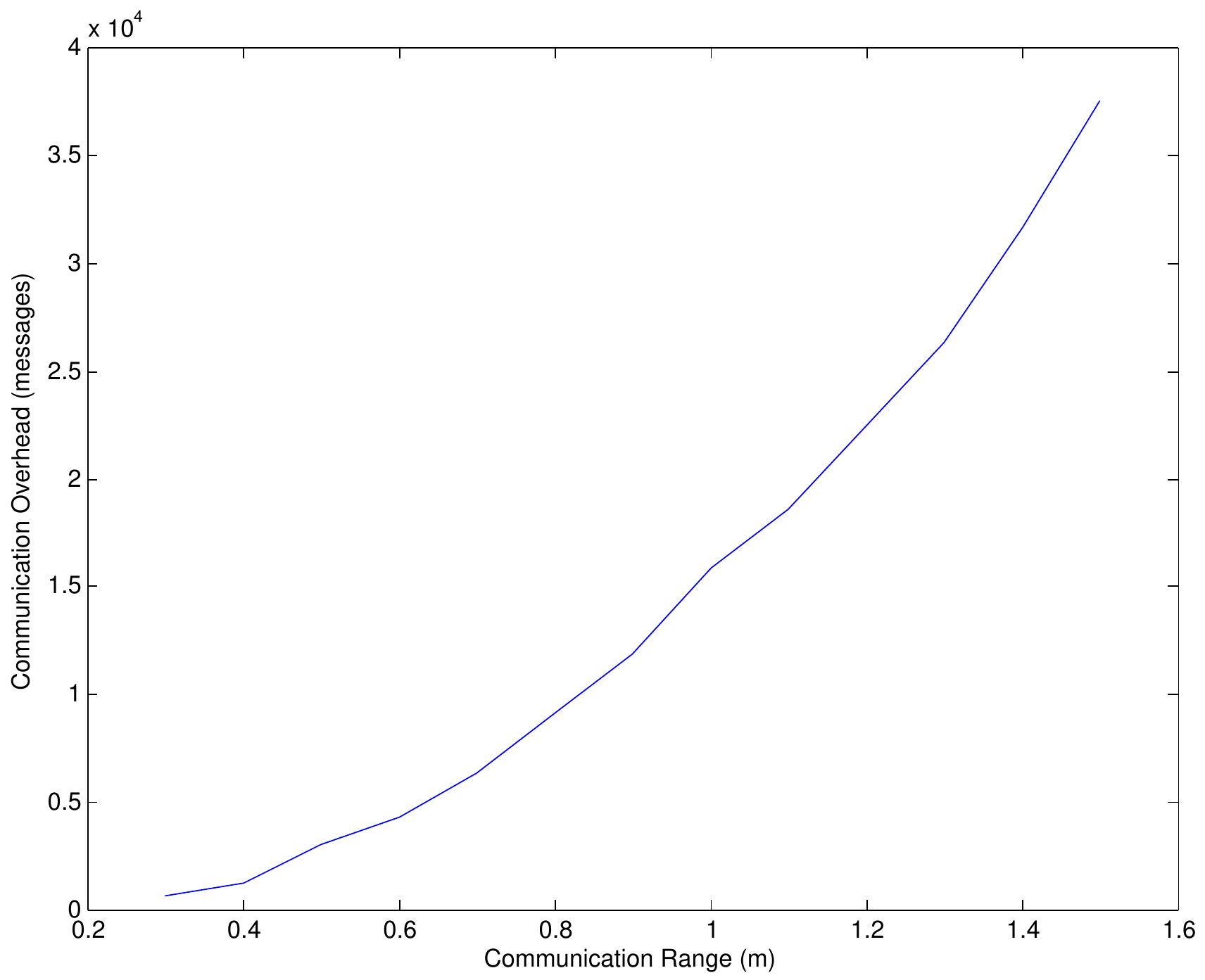}
        \caption{Communication overhead over different communication ranges of Algorithm \ref{distributed}.}
        \label{fig:messages}
    \end{subfigure}
    \begin{subfigure}[b]{0.9\columnwidth}
        \includegraphics[width=\textwidth]{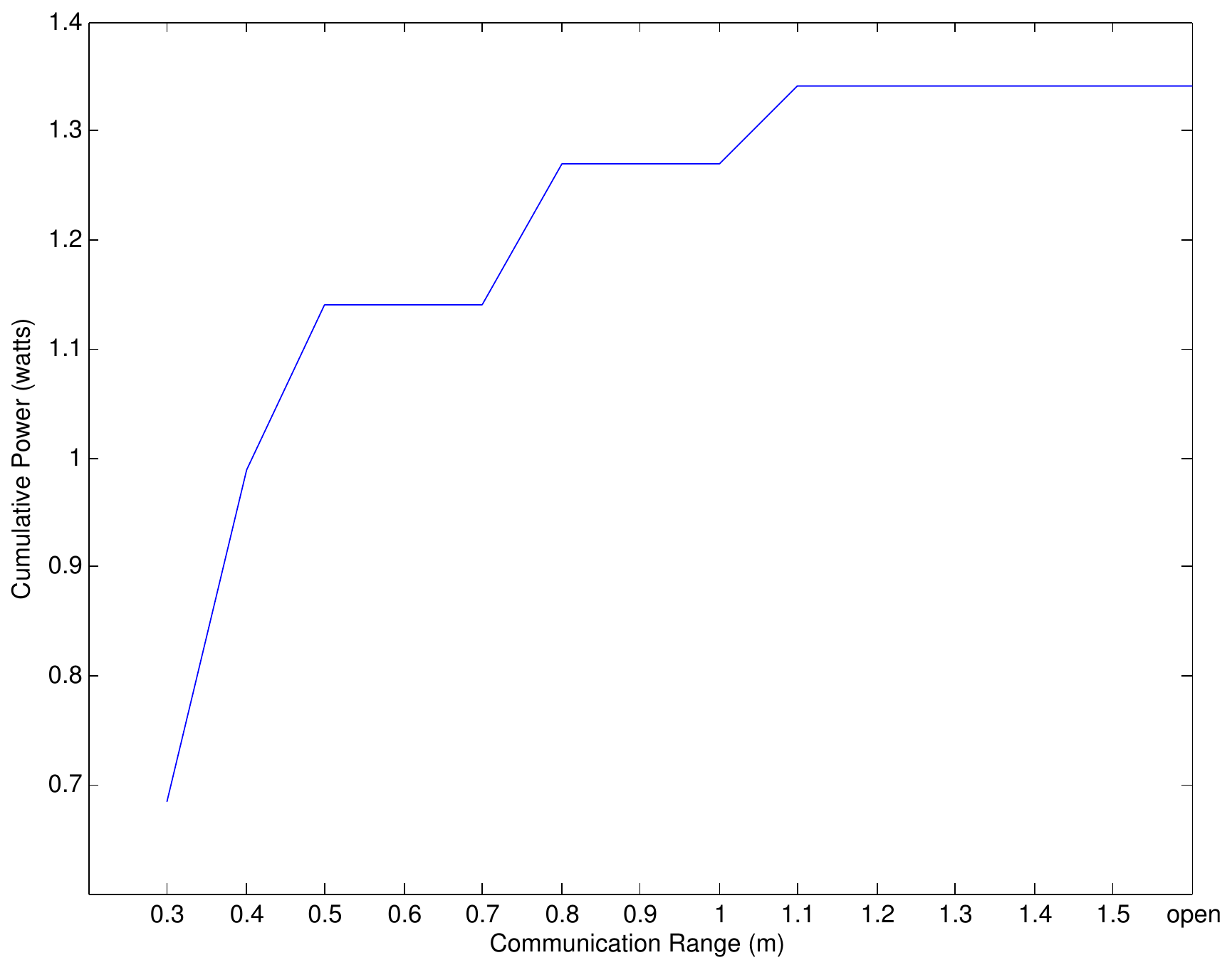}
        \caption{Cumulative power over different communication ranges at the $90th$ round of Algorithm \ref{distributed}.}
        \label{fig:CPperRange}
    \end{subfigure}
    ~ %add desired spacing between images, e. g. ~, \quad, \qquad, \hfill etc. 
    %(or a blank line to force the subfigure onto a new line)
    \begin{subfigure}[b]{0.9\columnwidth}
        \includegraphics[width=\textwidth]{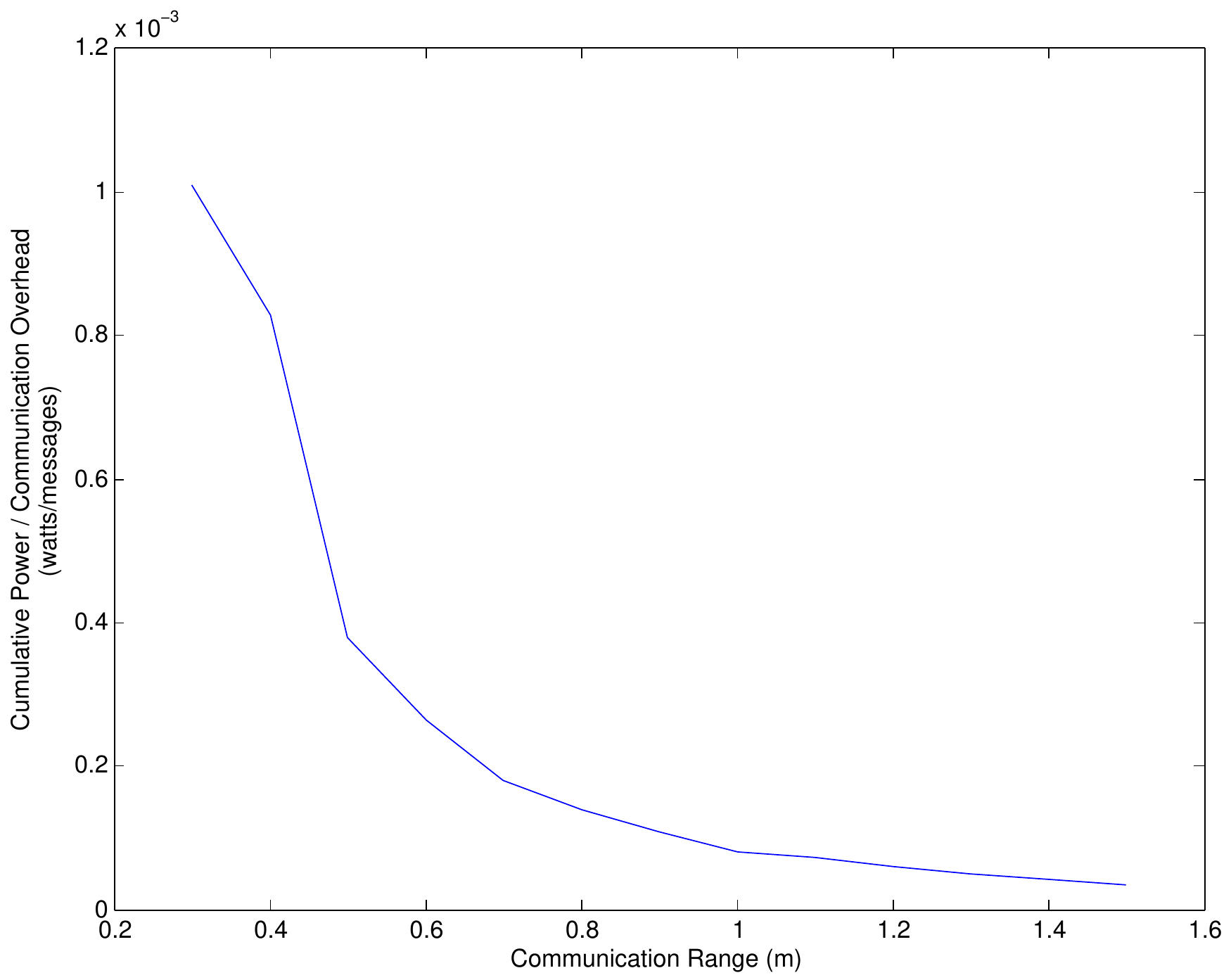}
        \caption{Ratio of cumulative power to communication overhead over different communication ranges.}
        \label{fig:ratio}
    \end{subfigure}
    \caption{Impact of communication range on different metrics.}\label{fig:}
\end{figure}

\textbf{Communication Overhead:}
Fig \ref{fig:messages} meters the number of messages that have been exchanged in the system during the running time of the distributed algorithm \ref{distributed} for ranges from $0.3$ to $1.5$. When the communication range of the chargers increases, the communication overhead rises. The effect of different communication ranges (from $0.3$ to $1.5$ and $open$) on the cumulative received power is depicted in Fig \ref{fig:CPperRange}. As we mentioned above a limited communication range ($1.1m-1.5m$) can achieve the performance of $open$. We observe that there is a trade-off between the communication overhead and the cumulative power that the nodes receive (Fig \ref{fig:ratio}). It is evident that the contribution of the messages drops for communication range bigger than $1m-1.2m$. Note that the communication overhead depends on the relative position of the chargers and the nodes. The chargers' placement for communication optimization is not considered in this study.

\begin{figure}[t]
    \centering  
    \begin{subfigure}[b]{\columnwidth}
        \includegraphics[width=\textwidth]{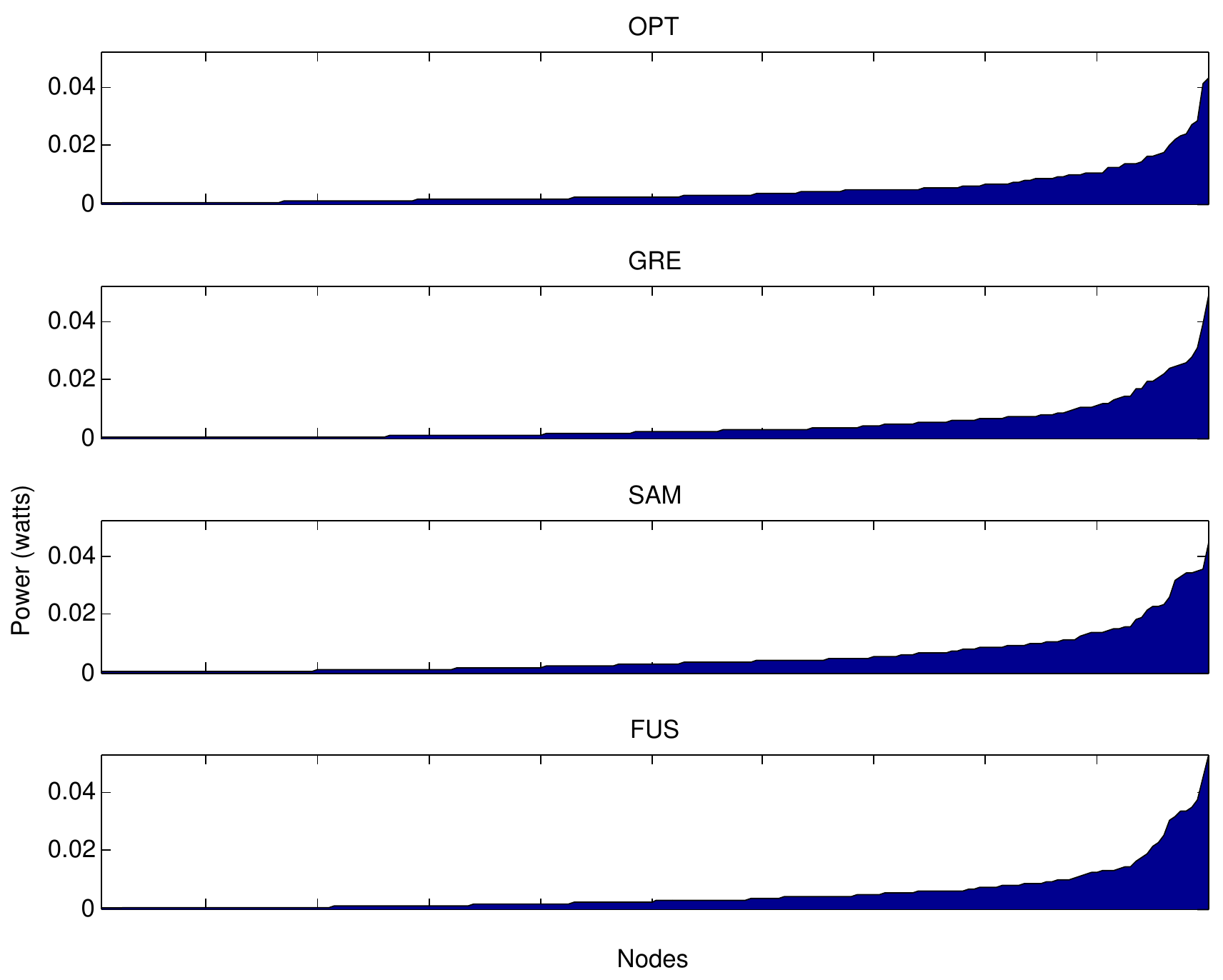}
        \captionsetup{justification=centering}
        %\caption{Power balance in general case}
        %\label{fig:pbgeneral}
    \end{subfigure}
    \caption{Power balance achieved by different algorithms.}\label{fig:pbgeneral}
\end{figure}

\textbf{Power Balance:}
We finally study the impact of our methods on the system in relation to power balance (the variation of  power among the various nodes).
The simulation results are shown in Fig \ref{fig:pbgeneral}, where we examine the performance of the algorithms in the general case. Obviously, there is room for improvement in this area, as our methods do not take power balance into consideration and we leave this as an open problem for future work.

\section{Conclusion and Future Work}\label{conclusion}
In this chapter, we studied the problem of finding the best configuration setup of the wireless power transmitters in a wireless system with respect to the maximization of the cumulative received power and we proposed heuristic algorithms to succeed in this goal. Finally, we evaluated the performance of the proposed algorithms through simulations, and provided numerical results to validate their efficiency. A main contribution of our work lies on the fact that, for the first time power maximization algorithms are given under the vector model which realistically addresses the superposition of energy fields.

In future work, we opt to improve the power balance of the algorithms in exchange of a small amount of the cumulative received power. We will also explore solutions with good approximation ratios to maximize the cumulative received power from the point of view of nodes or chargers deployment.

%\section*{References}

%\bibliography{sigproc}

\end{document}